%% file: preprint-improving-communication.tex
\documentclass{article}

\input{A02_utils/command_configs}

\input{A02_utils/packages}
\usepackage{arxiv}

\hypersetup{
    colorlinks=true, % Remove link borders
    linkcolor=blue, % Color for internal links
    urlcolor=blue, % Color for URLs (including DOIs)
    citecolor=blue, % Color for citations
    pdfborderstyle={/S/U/W 0}, % Remove border around links
    pdfborder={0 0 0} % Remove border entirely (alternative)
}
\title{Improving Communication of Changes in Model-Based Engineering with Model-Independent Change Descriptions}
\author{% 
    \href{https://orcid.org/0009-0009-4224-3530}{%
    \includegraphics[scale=0.06]{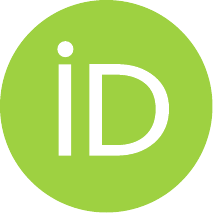}%
    \hspace{1mm}Philip Ochs} \\
	Institute of Information Security and Dependability\\
	Karlsruhe Institute of Technology\\
	Am Fasanengarten 5, Karlsruhe, 76131, Germany. \\
	\texttt{philip.ochs@kit.edu} \\
	\And
    \href{https://orcid.org/0009-0009-7786-4847}{
    \includegraphics[scale=0.06]{orcid.pdf}
    \hspace{1mm}Lars Gesmann} \\
	Institute of Product Engineering\\
	Karlsruhe Institute of Technology\\
	Kaiserstrasse 10, Karlsruhe, 76131, Germany. \\
	\texttt{lars.gesmann@kit.edu} \\
    \And
    \href{https://orcid.org/0000-0001-7652-6525}{%
    \includegraphics[scale=0.06]{orcid.pdf}%
    \hspace{1mm}Tobias Pett} \\
	Institute of Information Security and Dependability\\
	Karlsruhe Institute of Technology\\
	Am Fasanengarten 5, Karlsruhe, 76131, Germany. \\
	\texttt{tobias.pett@kit.edu} \\
    \And
    \href{https://orcid.org/0000-0002-7153-761X}{%
    \includegraphics[scale=0.06]{orcid.pdf}%
    \hspace{1mm}Ina Schaefer} \\
	Institute of Information Security and Dependability\\
	Karlsruhe Institute of Technology\\
	Am Fasanengarten 5, Karlsruhe, 76131, Germany. \\
	\texttt{ina.schaefer@kit.edu} \\
}
\renewcommand{\shorttitle}{Improving Communication of Changes in Model-Based Engineering}
\hypersetup{
pdftitle={Improving Communication of Changes in Model-Based Engineering with Model-Independent Change Descriptions},
pdfsubject={consistency preservation in model-based engineering},
pdfauthor={Philip Ochs, Lars Gesmann, Tobias Pett, Ina Schaefer},
pdfkeywords={MBSE, change description, change communication, delta modeling},
}
\begin{document}
\maketitle

%%============================================================
%% Extra Includes
%%============================================================
\input{A02_utils/acronyms}
\input{A02_utils/colors}
\input{A02_utils/environments}
\input{A02_utils/terms_and_defs}
\input{A02_utils/urls}

% \input{A01_meta/title}
% \input{A01_meta/sn_authors}
%%============================================================

%%============================================================
%% Abstract
%%============================================================
\begin{abstract}
	\input{01_chapters/abstract}
\end{abstract}
%%============================================================

%%============================================================
%% Keywords
%%============================================================
% keywords can be removed
%%============================================================

%%============================================================
%% Disclaimer
%%============================================================
\setlength{\fboxsep}{10pt} % Padding inside the box
\setlength{\fboxrule}{1pt} % Border thickness

\definecolor{disclaimerBorder}{RGB}{0, 100, 200} % Darker blue for border
\definecolor{disclaimerBg}{RGB}{230, 240, 255} % Light blue background

\fcolorbox{disclaimerBorder}{disclaimerBg}{%
    \parbox{\dimexpr\linewidth-2\fboxsep-2\fboxrule\relax}{%
        \textbf{Disclaimer} \\[6pt] % Headline with spacing
        This preprint presents work that has been submitted to the \textit{International Journal on Software and Systems Modeling (SoSyM)} for peer review. \\[4pt]
        Please note that the manuscript may undergo further revisions before final publication. \\[4pt]
        The current version is dated \today. Copyright may belong to Springer. \\[4pt]
        \textit{This preprint is shared under the terms of arXiv's default license.}
    }%
}
%%============================================================

%%============================================================
%% Contents
%%============================================================
\input{A01_meta/contents}

\section*{Acknowledgements}
\input{01_chapters/acknowledgements}
%%============================================================

%%============================================================
%% Bibliography
%%============================================================
% \bibliographystyle{abbrvnat}
% \bibliography{references}  

\printbibliography
%%============================================================

\end{document}

%% file: A02_utils/command_configs.tex
\DeclareTextFontCommand{\texthtt}{\ttfamily\hyphenchar\font=45\relax}

\ExplSyntaxOn
\NewExpandableDocumentCommand{\fpcompareTF}{mmm}
 {
  \fp_compare:nTF { #1 } { #2 } { #3 }
 }
\ExplSyntaxOff

%% file: A02_utils/packages.tex
\usepackage{amsmath,amsfonts}
\usepackage{lipsum}
\usepackage{booktabs}
\usepackage{multirow}
\usepackage[table]{xcolor}
\usepackage[nolist]{acronym}
\usepackage{listings} % source code listings
\usepackage[T1]{fontenc}
\usepackage{hyperref}
\usepackage{enumitem}
\usepackage{graphicx}
\usepackage{textcomp}
\usepackage{hyperref}
\usepackage{subcaption}
\usepackage{pdfpages}
\usepackage{todonotes}
\usepackage{svg}

\usepackage[strings]{underscore}

\usepackage{adjustbox}
\usepackage{array}

\usepackage[utf8]{inputenc} % allow utf-8 input
\usepackage{url}            % simple URL typesetting
\usepackage[style=numeric, doi=true, url=false, backend=biber]{biblatex}

%% file: A02_utils/acronyms.tex
% pattern \acrodef{id}[short]{long}

\acrodef{sge}[SGE]{System Generation Engineering}
%\acrodef{sge}[pmSGE]{process model of System Generation Engineering}
\acrodef{cv}[CV]{Carryover Variation}
\acrodef{av}[AV]{Attribute Variation}
\acrodef{pv}[PV]{Principle Variation}
\acrodef{uml}[UML]{Unified Modeling Language}
\acrodef{sysml}[SysML]{Systems Modeling Language}
\acrodef{json}[JSON]{JavaScript Object Notation}
\acrodef{ocl}[OCL]{Object Constraint Language}

\acrodef{bcs}[BCS]{Body Comfort System}
\acrodef{smm}[SMM]{State Machine Model}
\acrodef{cam}[CAM]{Component Architecture Model}
\acrodef{emf}[EMF]{Eclipse Modeling Framework}
\acrodef{mof}[MOF]{Meta Object Facility}

\acrodef{rq}[RQ]{research question}

%% file: A02_utils/colors.tex
% pattern \definecolor{id}{type}{value}
\definecolor{lightgreen}{HTML}{F5FFFA}
\definecolor{lightergreen}{HTML}{DAF5E8}
\definecolor{darkgreen}{HTML}{47916E}

%% KIT Green
\definecolor{kitGreen1}{HTML}{00876c}
\definecolor{kitGreen2}{HTML}{b7fff5}
\definecolor{kitGreen3}{HTML}{6fffec}
\definecolor{kitGreen4}{HTML}{27ffe2}
\definecolor{kitGreen5}{HTML}{007162}
\definecolor{kitGreen6}{HTML}{004b41}
%% =========================================
%% KIT Blue
\definecolor{kitBlue1}{HTML}{4664aa}
\definecolor{kitBlue2}{HTML}{d9dfef}
\definecolor{kitBlue3}{HTML}{b3c0df}
\definecolor{kitBlue4}{HTML}{8ca1d0}
\definecolor{kitBlue5}{HTML}{354b7f}
\definecolor{kitBlue6}{HTML}{233255}
%% =========================================
%% KIT Black
\definecolor{kitBlack1}{HTML}{404040}
\definecolor{kitBlack2}{HTML}{000000}
\definecolor{kitBlack3}{HTML}{7f7f7f}
\definecolor{kitBlack4}{HTML}{595959}
\definecolor{kitBlack5}{HTML}{262626}
\definecolor{kitBlack6}{HTML}{0d0d0d}
%% =========================================
%% KIT May Green
\definecolor{kitMayGreen1}{HTML}{77a200}
\definecolor{kitMayGreen2}{HTML}{e8f2d7}
\definecolor{kitMayGreen3}{HTML}{d2e4ae}
\definecolor{kitMayGreen4}{HTML}{bbd786}
\definecolor{kitMayGreen5}{HTML}{69882d}
\definecolor{kitMayGreen6}{HTML}{465b1e}
%% =========================================
%% KIT Purple
\definecolor{kitPurple1}{HTML}{a3107c}
\definecolor{kitPurple2}{HTML}{f9c3eb}
\definecolor{kitPurple3}{HTML}{f386d6}
\definecolor{kitPurple4}{HTML}{ed4ac2}
\definecolor{kitPurple5}{HTML}{7a0c5d}
\definecolor{kitPurple6}{HTML}{52083e}
%% =========================================
%% KIT Brown
\definecolor{kitBrown1}{HTML}{a7822e}
\definecolor{kitBrown2}{HTML}{df9b1b}
\definecolor{kitBrown3}{HTML}{f9ebd1}
\definecolor{kitBrown4}{HTML}{f4d7a2}
\definecolor{kitBrown5}{HTML}{eec474}
\definecolor{kitBrown6}{HTML}{6f4e0e}
%% =========================================
%% KIT Cyan
\definecolor{kitCyan1}{HTML}{079ede}
\definecolor{kitCyan2}{HTML}{d3ecf9}
\definecolor{kitCyan3}{HTML}{a7d9f3}
\definecolor{kitCyan4}{HTML}{7bc7ec}
\definecolor{kitCyan5}{HTML}{187aaa}
\definecolor{kitCyan6}{HTML}{105172}
%% =========================================
%% KIT Gray
\definecolor{kitGray1}{HTML}{d9d9d9}
\definecolor{kitGray2}{HTML}{c3c3c3}
\definecolor{kitGray3}{HTML}{a3a3a3}
\definecolor{kitGray4}{HTML}{6d6d6d}
\definecolor{kitGray5}{HTML}{363636}
\definecolor{kitGray6}{HTML}{161616}
%% =========================================
%% KIT Yellow
\definecolor{kitYellow1}{HTML}{fce500}
\definecolor{kitYellow2}{HTML}{fffacb}
\definecolor{kitYellow3}{HTML}{fff698}
\definecolor{kitYellow4}{HTML}{fff164}
\definecolor{kitYellow5}{HTML}{bdac00}
\definecolor{kitYellow6}{HTML}{7e7200}
%% =========================================
%% KIT White
\definecolor{kitWhite1}{HTML}{ffffff}
\definecolor{kitWhite2}{HTML}{f2f2f2}
\definecolor{kitWhite3}{HTML}{d9d9d9}
\definecolor{kitWhite4}{HTML}{bfbfbf}
\definecolor{kitWhite5}{HTML}{a6a6a6}
\definecolor{kitWhite6}{HTML}{7f7f7f}

%% file: A02_utils/environments.tex
\newlist{questions}{enumerate}{2}
\setlist[questions,1]{label=RQ\arabic*.,ref=RQ\arabic*}
\setlist[questions,2]{label=(\alph*),ref=\thequestionsi(\alph*)}

% quote environment with normal-sized font
\newenvironment{nquote}{\vspace{0.2cm}\begin{quote}\normalsize}{\end{quote}\vspace{0.2cm}}

\newcounter{finding}
\newenvironment{finding}
    {\refstepcounter{finding}\textbf{Finding F{query}}:}
    {}

\newcounter{query}
%\makeatletter
\newenvironment{query}
{\refstepcounter{finding}\textbf{Finding F{query}}:}
{}
%\makeatother
% Attempt to override the name autoref uses for the query environment
\newcommand{\queryautorefname}{Query}

%% file: A02_utils/terms_and_defs.tex
% pattern \newcommand{\id}{value}
% parameters \newcommand{\id}[no of parameters]{value with #1 ...}

\newcommand{\resq}[1]{\ac{rq}{#1}}

\newcommand{\attrdiskdiameter}{\texthtt{disk\-Diameter}}
\newcommand{\version}[1]{$\text{V}_{#1}$}
\newcommand{\product}[1]{$\text{P}_{#1}$}

\newcommand{\versionjump}[2]{\version{{#1}}$\;\rightarrow\;$\version{{#2}}}

\newcommand{\cpscommunicator}{\emph{cps-communicator}}

\newcommand{\model}[1]{M_{#1}}
\newcommand{\modelelementspace}[1]{E_{#1}}
\newcommand{\modelelement}[1]{e_{#1}}
\newcommand{\modelrelationspace}[1]{R_{#1}}
\newcommand{\modelrelation}[2]{r(#1, #2)}

\newcommand{\operation}[1]{\mathit{op}_{#1}}
\newcommand{\opaddelement}[1]{\mathtt{add}\ #1}
\newcommand{\opmodelement}[1]{\mathtt{mod}\ #1}
\newcommand{\opdelelement}[1]{\mathtt{del}\ #1}
\newcommand{\operationspace}[1]{\mathit{Op}_{#1}}

\newcommand{\modelop}[2]{\mathit{op}_{#1,#2}}
\newcommand{\modelopspace}[1]{\mathit{Op}_{#1}}

\newcommand{\sgevar}[1]{\mathit{var}_{#1}}
\newcommand{\sgetypecv}[1]{\mathtt{CV}\ #1}
\newcommand{\sgetypeav}[1]{\mathtt{AV}\ #1}
\newcommand{\sgetypepv}[1]{\mathtt{PV}\ #1}
\newcommand{\sgevarspace}[1]{\mathit{Var}_{#1}}

\newcommand{\mapadef}{\mathit{map}_a}
\newcommand{\mapa}[1]{\mapadef(#1)}
\newcommand{\mapcdef}{\mathit{map}_c}
\newcommand{\mapc}[1]{\mapcdef(#1)}

\newcommand{\category}[1]{\textsf{{#1}}} % sans-serif category notion
\newcommand{\codecategory}[1]{\texttt{C{#1}}}
\newcommand{\code}[2]{\codecategory{{#1}}\texttt{–{#2}}}

\newcolumntype{A}[2]{%
    >{\adjustbox{angle=#1,lap=\width-(#2)}\bgroup}%
    l%
    <{\egroup}%
}
\newcolumntype{C}[1]{>{\centering\arraybackslash}p{#1}} % center-aligned, fixed width table columns
\newcolumntype{R}[1]{>{\raggedleft\arraybackslash}p{#1}} % right-aligned, fixed width table columns

% cellcolor depending on numeric cell value (in [0, 100])
\newcommand{\relcellcolor}[5]{%
    \cellcolor{%
        #2!\fpeval{round(100*(#4/100),0)}!#1%
    }%
    {#3}{#4}{#5}%
}
\newcommand{\relgreen}[1]{\relcellcolor{white}{green}{(}{{#1}}{)}} % example for range from white to green

%% file: A02_utils/urls.tex
% pattern \urldef{\id}\url{https://...}

\urldef{\emf}\url{https://projects.eclipse.org/projects/modeling.emf.emf}
\urldef{\eclipse}\url{https://www.eclipse.org/downloads/packages/release/2024-12/r/eclipse-modeling-tools}
\urldef{\gradle}\url{https://gradle.org}
\urldef{\git}\url{https://git-scm.com}
\urldef{\java}\url{https://www.java.com}
\urldef{\papyrus}\url{https://eclipse.dev/papyrus}
\urldef{\xtext}\url{https://eclipse.dev/Xtext}
\urldef{\deltaecore}\url{https://github.com/chseidl/deltaecore}
\urldef{\bcs}\url{https://github.com/TUBS-ISF/BCS-Case-Study-Full}
\urldef{\mof}\url{https://www.omg.org/spec/MOF}
\urldef{\zenodo}\url{tbd.}

\urldef{\cpscomm}\url{https://github.com/KIT-TVA/cps-communicator}

\urldef{\teams}\url{https://www.microsoft.com/microsoft-teams}
\urldef{\miro}\url{https://miro.com}
\urldef{\atrain}\url{https://business-analytics.uni-graz.at/en/research/atrain/}
\urldef{\maxqda}\url{https://www.maxqda.com/de/produkte/maxqda}

%% file: 01_chapters/abstract.tex
In model-based engineering, inter-disciplinary teams collaborate through models, which change over time for purposes of system development, what makes the proper description of such changes crucial for engineers. However, any change made by the engineer of one discipline will be difficult to understand by the engineers of other disciplines. To overcome this limitation, model-independent change descriptions can be derived instead, which preserve semantics of the changes and do not require model-specific knowledge. The two opposing approaches here are to describe changes using either informal language or formal notions of change. While informal language lacks objectivity and standardisation, formal notions of change lack human interpretability, and thus offering no support for inter-disciplinary communication. In this paper, we propose functions to map formally specified changes, represented in the approach of delta modelling, to change descriptions in model-independent language. In an exhaustive mixed-methods evaluation, we bridge the gap between the theoretical and the practical representation of changes. We quantitatively assess technical feasibility, with an implementation framework, and technical applicability, along a case study; and we qualitatively assess plausibility, practical applicability, and extensibility, in a user study. Our work shows a promising starting point for automated, model-independent description of changes in model-based engineering projects.

% In multi-disciplinary engineering, the communication of changes over time is critical for preserving consistency within the project, but highly model-specific, and thus difficult to understand by engineers outside the domain in which the change was initiated. 

%Expert-based efforts to derive model-independent change descriptions are either not feasible in large engineering projects with possibly many occurring changes, or subjective to the human translator. 

%In this paper, we approach the challenge of communicating changes across domains in large engineering projects with a concept to translate model-specific changes into model-independent change descriptions, while preserving semantic information of the change itself. Therefore, we propose functions to map formally specified changes, represented in the approach of delta modelling, to change descriptions in inter-disciplinary language. In an exhaustive mixed-methods evaluation, we bridge the gap between the theoretical and the practical representation of changes. We quantitatively assess technical feasibility, with an implementation framework, and technical applicability, along a case study; and we qualitatively assess plausibility, practical applicability, and extensibility, in a user study. Our work shows a promising starting point for advanced description of changes in inter-disciplinary, model-based engineering projects.

%% file: A01_meta/contents.tex
\input{01_chapters/introduction}

\input{01_chapters/background}

% %\input{01_chapters/runningexample}
\input{01_chapters/concept}
\input{01_chapters/evaluation}

% %\input{01_chapters/extensions_and_limitations}
\input{01_chapters/related_work}
\input{01_chapters/conclusion}

%% file: 01_chapters/introduction.tex
\section{Introduction}

Model-based cyber-physical engineering \cite{technicaloperationsinternationalcouncilonsystemsengineeringincoseINCOSESystemsEngineering2007} requires the pooling of expertise from multiple disciplinary backgrounds. Teams of engineers work in coordinated collaboration on models but also independently make changes to the models for purposes of system development. Hence, both the development and also the collaboration between developers depend crucially upon appropriately describing model changes over time. Inconsistencies between the models are hard to detect manually but lead to significant drawbacks and introduce high financial risks for the manufacturers \cite{spanoudakisInconsistencyManagementSoftware2001}. %For example, when change descriptions are inaccurate or incomprehensible, further analyses such as risk assessments are severely hindered or even become impossible.

To preserve consistency, a change to a model must be described to other engineers across the domains involved so that they are aware of the change and can initiate subsequent changes for consistency preservation to affected models. However, a concise description of a model change is highly domain-specific, and thus hard to understand by engineers outside the domain in which the change was initiated. This opens the challenge of describing highly domain-specific model changes in order to preserve consistency of the models involved throughout the engineering project.

Informal languages for the description of changes, such as natural language used in code commit messages, are well interpretable by engineers but are usually subjective and lack standardisation \cite{kamsties_detecting_2001}. Furthermore, concrete change descriptions in informal languages must be set up manually and so lack crucial parts for automated processing. Consequently, informal languages are infeasible in the sorts of large projects typical of cyber-physical engineering. On the other hand, model-specific change descriptions in formal languages can be processed computationally but are neither supportive for human interpretability \cite{knight_challenges_1998} nor for interdisciplinary communication, as necessary in cyber-physical engineering processes.

% In a model-based engineering process of a cyber-physical system \cite{technicaloperationsinternationalcouncilonsystemsengineeringincoseINCOSESystemsEngineering2007}, inconsistencies between the models are hard to detect manually but lead to significant drawbacks and harbour high financial risks for the manufacturers \cite{spanoudakisInconsistencyManagementSoftware2001}. To preserve consistency, a change to a model must be communicated and propagated to other engineers across the domains involved, so that they are aware of the change and can initiate subsequent changes for consistency preservation to affected models. However, a concise description of a model change is highly domain-specific, and thus hard to understand by engineers outside the domain in which the change was initiated. This opens the challenge of communicating highly domain-specific model changes in order to preserve consistency of the models involved throughout the engineering project.

%Informal languages for the description of changes, such as natural language used in code commit messages, are well interpretable by engineers but are usually subjective and miss standardisation. Furthermore, concrete change descriptions in informal languages must be set up manually and so lack crucial parts for automated processing. In turn, model-specific change descriptions in \emph{formal} languages can be processed computationally but are neither supportive for human interpretability nor for interdisciplinary communication, as necessary in cyber-physical engineering processes.

In this paper, we approach the challenge of describing highly domain-specific model changes in multi-disciplinary engineering, by combining formal, computation-based languages with informal, human-interpretable languages. For that, we propose mapping functions to translate formalised model changes into model-independent, human-interpretable \emph{change descriptors}. As a representation of formal languages of change, we use the concept of delta modelling \cite{schaeferVariabilityModellingModelDriven2010}, originating in the field of software product lines, for expressing changes over time on arbitrary models. As a representation of informal languages of change, we use the description model of \ac{sge} \cite{albersProductGenerationDevelopment2015, albersModelSGESystem2022}, originating in the field of mechanical engineering, providing classification and metrics of change. In an extensive, threefold mixed-methods evaluation, we bridge the gap between formal and informal description of change. We quantitatively assess (1) technical feasibility and (2) technical applicability of our approach, by implementing a generic framework for change descriptors, and applying it to \acp{cam} and \acp{smm} of the real-world \ac{bcs} case study \cite{lityDeltaorientedSoftwareProduct2012} as subject system. We qualitatively assess (3) plausibility, practical applicability, and extensibility of our proposed approach, by conducting a study with twelve experts, from the research field of the description model of \ac{sge}, containing a semi-structured interview.
%Furthermore, we enable our concept to be extensible for further change analysis dimensions as well as the ability to deal with uncertainties in the translation to change descriptors. Finally, we evaluate the technical feasibility, applicability and plausibility of our proposed approach on component architecture models and state machine models along an adaption of the real-world \ac{bcs} case study \cite{lity_delta-oriented_2012}.

Our contributions advance the interdisciplinary communication and propagation of model changes within model-based engineering projects, by enabling automated, instead of manual description and analysis of evolution. This again simplifies preservation of the project's consistency, thus lowering development risk for manufacturers.

The remainder of this paper is structured as follows: In \autoref{sec:background}, we introduce fundamental concepts to our work as well as a running example. In \autoref{sec:concept}, we present the concept of our description approach in detail, which we exhaustively evaluate in \autoref{sec:evaluation}. In \autoref{sec:related_work}, we discuss existing literature, related to the description of changes. Finally, in \autoref{sec:conclusion}, we summarise our work and outlook to future work.

%% file: 01_chapters/background.tex
\section{Background}
\label{sec:background}

This section covers concepts which serve as a foundation of our work. We first introduce delta modelling \cite{schaeferVariabilityModellingModelDriven2010} as a representative of a formal, computation-based change description language, together with its context of software product lines. Then we introduce the description model of \ac{sge} \cite{albersModelSGESystem2022} as a representative of an informal, human-interpretable language for change.% Last, we introduce a running example on which we illustrate our work throughout this paper.

\subsection{Software Product Lines \& Delta Modelling}

In software product lines, individual product variants share common properties, but differ in their functionality, introducing \emph{variability in space} \cite{clementsSoftwareProductLines2002}. The definition of possible functionalities, called \emph{features}, and the theoretical composition of them into individual \emph{configurations}, for example for a specific customer, are given in the problem space of a product line. In turn, in the solution space, concrete realisation artefacts for a configuration are derived. The realisation artefacts comprehend, in context of this work, models, for example from the \ac{sysml} standard \cite{technicaloperationsinternationalcouncilonsystemsengineeringincoseINCOSESystemsEngineering2007}. Thus, within a single model, we refer to a \emph{model variant} as a realisation artefact of a concrete configuration.

Delta Modelling, proposed by Schaefer \cite{schaeferVariabilityModellingModelDriven2010}, is an approach to derive model variants in a software product line. From a core model variant, individual model variants are derived by applying a combination of predefined \emph{deltas} to this core model variant. A delta itself consists of a set of \emph{delta operations}, which represent atomic operations, such as additions, modifications, and deletions to a model. A \emph{delta model} defines all applicable deltas to a model, together with their application conditions, based on selected and deselected features in a configuration. This means that, for a concrete configuration, it is known which deltas have to be applied to the core model variant in order to get the desired model variant. Thus, a delta model is a solution space representation of a product line's variability.

To compose deltas from delta operations, the \emph{delta dialect} of a model defines a set of delta operations applicable in the model. Specifically, delta operations change either modelling elements, or relations over modelling elements, by adding, modifying, or deleting them. For example, the delta dialect of a \ac{sysml} class model could contain delta operations for adding a class, or for modifying cardinalities in an relation between two classes. On the instance level in this example, the delta operations would add a specific instance of the class, or would modify cardinalities of a specific instance of the relation between two specific class instances.

To tackle growing size and complexity, and ease analysis for delta models, refactoring techniques concern structural properties of deltas, such as monotonicity \cite{damianiRefactoringDeltaOrientedProduct2016}. Monotonicity in context of delta modelling describes how a delta changes a model, to which the delta is applied, either increasingly (adding elements) or decreasingly (deleting elements). Specifically of interest in our work, as edge case consideration, are \emph{strictly-decreasing monotonic deltas}, which only contain delta operations for deletion of modelling elements. This means that, after applying a strictly-decreasing monotonic delta, a model only contains modelling elements which were existing before the delta was applied, \emph{and} which were not modified by the applied delta.

\subsection{\acf{sge}}

The description model of \acf{sge}, proposed by Albers et al. \cite{albersProductGenerationDevelopment2015, albersModelSGESystem2022}, originating in mechanical engineering, is an approach to describe and analyse the development of cyber-physical systems in interdisciplinary, model-based engineering. While it was tailored to mechanical systems initially \cite{albersProductGenerationDevelopment2015}, the authors later generalised their approach, to be applicable to cyber-physical systems in model-based engineering \cite{albersModelSGESystem2022}.

The description model of \ac{sge} is predicated on the hypothesis that a new system is always built upon subsystems, contained in a \emph{reference system}, varied and integrated into the new system. There exist three types of variations to a subsystem: With (1) a \ac{cv}, a subsystem is taken almost without change, only with adjusted interfaces, integrated into the new system. With (2) an \ac{av}, a subsystem is changed in the properties of its contained elements, but maintaining structural and functional relations. In (3) a \ac{pv}, a subsystem is changed in the relations of its contained elements, thus in its functional principle. In combination, a new system hence is described by the union of all variations applied to the subsystems of the reference system, classified by the three variation types: The set of subsystems varied by \acp{cv}, the set of subsystems varied by \acp{av}, and the set of subsystems varied by \acp{pv}.

These three sets of subsystems, for \acp{cv}, \acp{av}, and \acp{pv}, are then used in product development analyses. Specifically, their cardinalities are set into relation to each other, thus the numerical shares of each variation type in the new system, so called \emph{variation shares}, are calculated. Variation shares are indicators for assessing innovation potential and development risks. For example, a high share of \acp{cv} indicates reduced development risk, because mostly known technologies are used. In turn, a high share of \acp{av} and \acp{pv} indicates the opposite: increased development risk because of more technical novelty.

%% file: 01_chapters/concept.tex
%\section{Describing Changes}
\section{Translating Model-Specific Changes into Model-Independent Change Descriptors}
\label{sec:concept}

We formalise the specification of model changes in a model-based engineering process so as to enable an automated description of changes. We use delta modelling \cite{schaeferVariabilityModellingModelDriven2010}, because it can be applied on an arbitrary model, as long as a meta-model is defined for it. This versatility of delta modelling is crucial in multi-domain engineering projects with many different models involved. However, as delta modelling was initially designed in the context of variability in space, we adapt it to our context of variability \textit{in time} (\autoref{sec:concept:dm_for_evolution}).

We use formalised changes, from our adaption of delta modelling, and lift them up to model-independent change descriptions, understandable outside the domain of the underlying model, by defining two mapping functions: (1) \emph{atomic mappings} specify the translation of delta operations into change descriptors, explained in \autoref{sec:concept:atomic_mappings}, and (2) \emph{composed mappings} specify the translation of deltas into aggregated change descriptors, explained in \autoref{sec:concept:composed_mappings}. Throughout the sections, we give application examples of change descriptors and aggregated change descriptors within the description model of \ac{sge} as model-independent language, and illustrations along a running example.

\subsection{Preliminaries: Delta Modelling for Variability in Time}
\label{sec:concept:dm_for_evolution}

Delta modelling \cite{schaeferVariabilityModellingModelDriven2010} represents an approach to derive model variants in product lines. Thus, a delta initially is specified and applied in context of variability in space. In contrast, the context of our work is variability in time (evolution). However, we can adapt concepts from delta modelling to be used in our context, namely (1) deltas as unit of versioning of a model, (2) delta operations as atomic changes to a model, and (3) delta dialects as specification of possible atomic changes for a model. We elaborate on these three parts in the following, and illustrate along our running example.

We use a delta to represent changes between two (consecutive) versions of a model. This means, starting from an initial version of a model, with a delta we formally describe an ordered sequence of change operations applied to a model, to then represent its next version.

Delta operations contained in a delta are applied to modelling elements of a model, thus either add new, modify, or delete existing modelling elements. For example, delta operations directly represent atomic actions, done by a developer changing a model.

For a specific model, we define possible delta operations, which are allowed to be applied to model instances, with an associated delta dialect. The delta dialect of a model does not need to be complete, i.e. not all modelling elements must necessarily be changeable by applying delta operations. For example, an identification attribute of a modelling element is typically designed to be immutable, thus no possibility to modify such an attribute is defined in the associated delta dialect.

\paragraph{Running Example}

\begin{figure}
    \centering
    %\includesvg[width=0.49\linewidth]{02_figures/running_example_delta_dialect.svg}
    \includegraphics[width=0.49\linewidth]{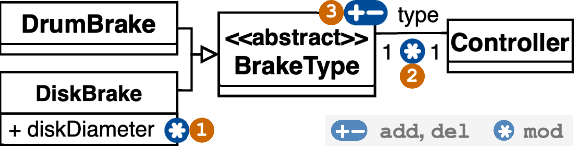}
    \caption{\acs{uml} class model of the running example representing an electro mechanical brake system, together with its delta dialect (represented with coloured symbols).}
    \label{fig:running_example_delta_dialect}
\end{figure}

Our running example represents a part of the controller software of an electro-mechanical brake system, thus picking up aspects of multi-domain engineering; software engineering and mechanical engineering specifically. The controller software is modelled as a \ac{uml} class model, as shown in \autoref{fig:running_example_delta_dialect} (without coloured symbols). Briefly, the controller is known to one brake type, either a drum brake or a disk brake; however, both these brake types share attributes (not modelled here explicitly) but each also has distinct, unshared attributes, such as the brake disk diameter of the disk brake.

The delta dialect for the \ac{uml} class model of our running example is marked with coloured symbols in \autoref{fig:running_example_delta_dialect}, and specifies the following delta operations: (1) the modification of the diameter of the disk brake, (2) the replacement of the actual brake type instance the controller is known to and (3) the addition and deletion of a brake type instance.

\subsection{Describing Delta Operations: Atomic Mappings}
\label{sec:concept:atomic_mappings}

To preserve semantic information of each (atomic) change operation while decoupling it from model-specific language, we propose atomic mappings. An atomic mapping is used to specify how a delta operation, defined in the delta dialect of a model, is translated into a change descriptor, defined in the model-independent language.

With the description model of \ac{sge} as model-independent language, we use a variation type (\ac{av}, \ac{pv}) as change descriptor. This means that, by applying an atomic mapping to an applied delta operation within a model, the delta operation is mapped to either an \ac{av}, which describes a change of one of the model's properties, or to a \ac{pv}, which describes a change of the model's functioning principle. We omit using \acp{cv}, because, according to Albers et al. \cite{albersModelSGESystem2022}, they describe minor changes to the interface of a carried-over modelling element ("carryover"), which, again, we understand as a change of a modelling element's properties.

%With the description model of \ac{sge} as model-independent language, we specify a change descriptor as a \emph{variation} with an associated \emph{variation type}. Thus, by applying an atomic mapping to an applied delta operation within a model, the delta operation is mapped to either an \ac{av}, which describes a change of one of the model's properties, or to a \ac{pv}, which describes a change of the model's functioning principle. We omit using \acp{cv}, because, according to Albers et al. \cite{albersProposingGeneralizedDescription2020}, they either describe not an actual change ("carryover"), or describe minor changes to the interface of a carried-over modelling element, which we, in turn, understand again as a change of a modelling element's properties.

%The technical realisation of the atomic mapping function is technology-independent: For example, declarative approaches can be used to specify which delta operations map to which variation types, or machine learning models can be trained on an expert-based set of classified delta operations.

\paragraph{Running Example}
For the \ac{uml} class model in our running example, we formalised a delta dialect with three delta operations, shown in \autoref{fig:running_example_delta_dialect}. We now specify the atomic mappings for the three delta operations as follows: (1) The modification of the diameter of the disk brake maps to an \ac{av}, because this delta operation concerns a (visible) attribute of the brake disk, and does not address any functioning principle of the brake system's \ac{uml} class model. (2) The modification of the actual brake type instance, the controller is known to, maps to a \ac{pv}, because this delta operation concerns the principle of how the brake system works physically. (3) The addition and deletion of a brake type instance are delta operations which are necessary to apply the delta operation for the replacement of the brake type instance (delta operation (2)); for example, first a new brake type instance is added, then the reference to the brake type of the controller is modified, and finally the former brake type instance is (optionally) deleted. We argue that the delta operations for adding and deleting a brake type instance are \acp{av} to this \ac{uml} class model, because we understand brake type instances as properties of the described brake system, and because the actual change of the brake system's function is represented by the modification of the controller's reference to a brake type instance.

\subsection{Describing Deltas: Composed Mappings}
\label{sec:concept:composed_mappings}

To compose change descriptions from an atomic level, covered by atomic mappings, we propose composed mappings. A composed mapping specifies how the impact of an applied delta to an underlying model is described model-independently, with an aggregated change descriptor. As we understand a delta as the \emph{model-specific} change description between two versions of an underlying model, an aggregated change descriptor thus represents the \emph{model-independent} change description between two versions of an underlying model.

Technically, we require the applied delta itself, the results of the applied atomic mappings for the delta, and the initial version of the model, to which the delta is applied, as inputs when applying a composed mapping; and we expect a string as output, which represents the concrete aggregated change descriptor. This way, composed mappings are versatile in their specification, and are enabled to aggregate results of applied atomic mappings as well as to work directly on the applied delta itself.

\newcommand{\variable}[1]{\texttt{<{#1}>}}
With the description model of \ac{sge} as model-independent language, we model an aggregated change descriptor as metric for development risk, taken by the manufacturer by applying a delta to an underlying model. This aligns to the understanding of \cite{albersProductGenerationDevelopment2015} on the central purpose of the description model of \ac{sge}. We realise this metric for development risk by combining two individual metrics, (1) about the extent to which a model have been changed by applying a delta, and (2) about the proportion of variation types which describe the contained delta operations within the delta. We realise (1) with the \emph{share of changed modelling elements}, by calculating the ratio between the number of changed modelling elements and the total number of modelling elements in a model, after applying the delta. We understand a modelling element to be changed iff (a) it is an entity instance and was either added or modified in one of its attributes or (b) it is an relation instance over entity instances and was either added or modified in one of its entity associations. We realise (2) with the proportion of variation types, by aggregating the results of the applied atomic mappings and counting the number of \acp{av} and \acp{pv}. In the resulting aggregated change descriptor, we concatenate both individual metrics into a string of format "\variable{x}\% changed modelling context, thereof \variable{y}\% \acp{av} and \variable{z}\% \acp{pv}", where \variable{x} is a variable for the share of changed modelling elements, \variable{y} is the number of \acp{av}, and \variable{z} is the number of \acp{pv}.

%\subsubsection{Remarks on Deletion Operations}
\paragraph{Remarks on Strictly-Decreasing Monotonic Deltas}
Strictly-decreasing monotonic deltas \cite{damianiRefactoringDeltaOrientedProduct2016} contain only delta operations for deletion of modelling elements. As the share of changed modelling elements is calculated on changed modelling elements \emph{after} a delta is applied, none of the deleted modelling elements are counted in, because they are no longer in the model itself. This results in a share of changed modelling elements of 0\%, which may sound counter-intuitive. However, we argue that, from perspective of the successive model version implied by the application of a strictly-decreasing monotonic delta, all modelling elements existing are fully carried over (\acp{cv}) from the previous version of the model. This stays in line with the understanding of Albers et al. \cite{albersProductGenerationDevelopment2015}, where new versions are always developed only by reuse (\acp{cv}) and adaption (\acp{av} and \acp{pv}), thus ignoring non-reused or non-adapted modelling elements.
%However, our definition of the share of changed modelling elements implies particularly that a delta consisting of deletion operations only is assessed as a change with a share of changed modelling elements of 0\%. We argue that, from perspective of the successive model version implied by the application of such a delta, all modelling elements existing are fully carried over (\acp{cv}) from the previous version of the model. This stays in line with the understanding of Albers et al. \cite{albersProductGenerationDevelopment2015} where new versions are always developed only by reuse (\acp{cv}) and adaption (\acp{av} and \acp{pv}), thus ignoring non-reused or non-adapted elements.

%\subsubsection{Remarks on ...... }
%\todo[inline]{necessary? NO}
%Although deletion operations do not directly affect the share of changed modelling elements, they are still important for semantically describing a change between two versions of a model nevertheless. Thus, for communicating a change of a model to other domains in the engineering project, we see both information, (1) the set of variations within the process model of \ac{sge} as results of atomic mappings, as well as (2) the share of changed modelling elements, equally mandatory.

\begin{figure*}[h]
    \centering
    %\includesvg[width=1\textwidth]{02_figures/running_example_evolution_1_2.svg}
    \includegraphics[width=1\textwidth]{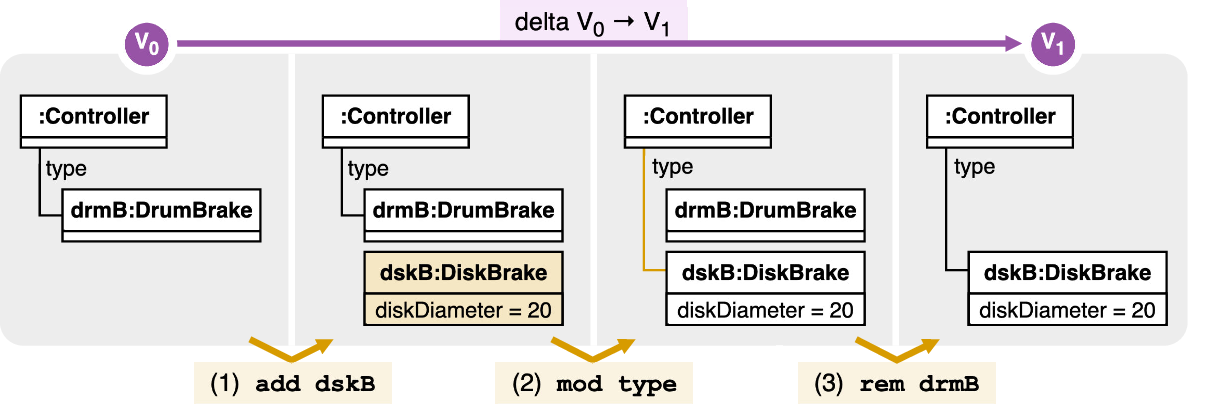}
    \caption{Evol. of the running example's \ac{uml} object model, version \version{0} to version \version{1}.}
    \label{fig:running_example_evolution_1_2}
\end{figure*}

\begin{figure}[h]
    \centering
    %\includesvg[width=0.5\linewidth]{02_figures/running_example_evolution_2_3.svg}
    \includegraphics[width=0.5\linewidth]{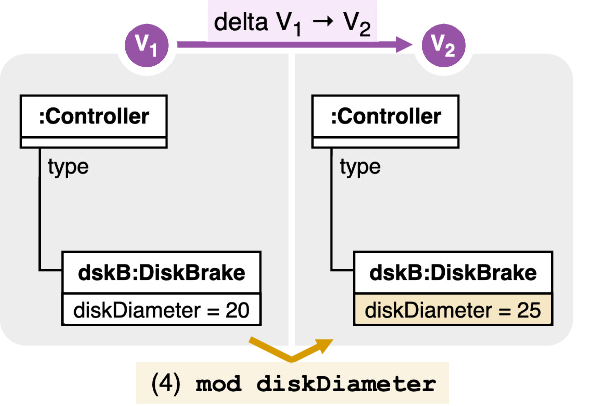}
    \caption{Evol. of the running example as \ac{uml} object model, version \version{1} to version \version{2}.}
    \label{fig:running_example_evolution_2_3}
\end{figure}

\paragraph{Running Example}
We illustrate an evolution scenario of our running example with three versions \version{0}, \version{1}, \version{2}. \autoref{fig:running_example_evolution_1_2} and \autoref{fig:running_example_evolution_2_3} show the evolution of \ac{uml} \emph{object} models, instantiated from the \ac{uml} class model introduced in \autoref{fig:running_example_delta_dialect}, with deltas coloured in purple, and delta operations coloured in orange. 

We assume an initial version \version{0} of our running example with a drum brake. The delta from version \version{0} to version \version{1} specifies a replacement from a drum brake to a disk brake, thus consisting of (1) an addition of a disk brake instance, (2) a modification of the brake type reference (to the new disk brake) of the controller and (3) a deletion of the old drum brake instance. We apply our atomic mapping specifications from \autoref{sec:concept:atomic_mappings} to the applied delta operations and map (1) the addition of the disk brake instance to an \ac{av}, (2) the modification of the brake type reference to a \ac{pv}, and (3) the deletion of the old drum brake instance, again, to an \ac{av}. Regarding the composed mapping specification, we directly calculate the proportion of variation types, from the results of the applied atomic mappings, which is 67\% \acp{av} and 33\% \acp{pv}. We then calculate the share of changed modelling elements: there are in total three modelling elements after applying the delta, namely the controller instance, the disk brake type instance and the relation between the controller and the brake type. Out of these three modelling elements, two have been changed: The disk brake type instance was added (delta operation (1)) and the reference from the controller instance to the brake type instance was modified (delta operation (2)). Therefore, we compute the share of changed modelling elements to be 67\%. The resulting aggregated change descriptor for the delta from version \version{0} to version \version{1} is "67\% changed modelling context, thereof 67\% \acp{av} and 33\% \acp{pv}".

In a further delta from version \version{1} to version \version{2}, a modification of the disk diameter of the disk brake instance is made. This applied delta operation is mapped to an \ac{av}, according to the atomic mapping. The resulting proportion of variation types thus is 100\% \acp{av} and 0\% \acp{pv}. The disk brake instance is the only modelling element in version \version{2} which has been changed, thus the share of changed modelling elements is 33\%.
Thus, the resulting change descriptor for the delta from version \version{1} to version \version{2} is "33\% changed modelling context, thereof 100\% \acp{av} and 0\% \acp{pv}".

%% file: 01_chapters/evaluation.tex
\section{Evaluation}
\label{sec:evaluation}

We present a mixed-methods evaluation \cite{johnsonDefinitionMixedMethods2007} of our approach for model-independent change descriptions. We quantitatively address \emph{technical feasibility} and \emph{technical applicability}, and qualitatively address \emph{plausibility}, \emph{practical applicability} and \emph{extensibility}. While delta modelling, as conceptual side of our approach, was evaluated for conceptual soundness \cite{schaeferVariabilityModellingModelDriven2010, clarkeAbstractDeltaModeling2010}, the description model of \ac{sge}, as practical side, was evaluated for usability in industry \cite{albersProductGenerationDevelopment2015, albersModelSGESystem2022, albersEntwicklungNachhaltigerSysteme2025}. Hence, our evaluation bridges the gap between the conceptual and the practical side. Particularly, we aim to answer the following \acp{rq}:

\begin{itemize}[leftmargin=28pt]
    \item[\textbf{\resq{1}}] We quantitatively evaluate \emph{technical feasibility} by answering the question: \textbf{How can model-specific delta operations and deltas be automatically translated into change descriptors and aggregated change descriptors?} We implement our approach in the Java programming language, to assess how definitions and functionalities of atomic and composed mappings can be abstracted from concrete applications, into a generalised framework. The implementation serves as a foundation for automated application of atomic and composed mappings.

    % OLD ---------
    %\item[\textbf{\resq{1}}] \textbf{[Technical Feasibility] How can model-specific deltas and delta operations be automatically translated to abstract change assessments and abstract change descriptors?} We implement our approach in the Java programming language and discuss how we achieve a generalised framework. The implementation serves as a foundation of automated application of atomic and composed mappings.
    %To show the technical feasibility of our approach, we describe how we implemented the composed mappings and atomic mappings into the Java programming language so that they can be executed automatically on arbitrary models and change description languages, for example \acp{cam} and \acp{smm} and the description model of \ac{sge}.
    
    \item[\textbf{\resq{2}}] We quantitatively evaluate \emph{technical applicability} by answering the question: \textbf{How can atomic mappings and composed mappings be applied to existing product evolution scenarios?} We use our implementation for the application to existing product evolution scenarios of our subject system, adapted from the automotive real-world \acf{bcs} case study \cite{lityDeltaorientedSoftwareProduct2012}, consisting of deltas and delta operations. We assess how delta operations can be mapped to change descriptors by atomic mappings, and how deltas can be mapped to aggregated change descriptors by composed mappings, within the description model of \ac{sge}.
    
    % OLD ---------
    %\item[\textbf{\resq{2}}] \textbf{[Technical Applicability] How can atomic mappings and composed mappings be applied to existing product evolution scenarios?} We apply our implementation to existing product evolution scenarios, adapted from the automotive real-world \acf{bcs} case study \cite{lity_delta-oriented_2012}, consisting of deltas and delta operations. For the results, we expect the deltas mapped to change assessments by composed mappings and the delta operations mapped to change descriptors by atomic mappings. We present and discuss the results in a quantitative manner.
    %We use our implemented composed mappings and atomic mappings from \resq{1} to analyse the necessary technical steps for applying our approach to the automotive real-world \acf{bcs} case study \cite{lity_delta-oriented_2012}. We translate given deltas and delta operations for \acp{cam} and \acp{smm} into the description model of \ac{sge}.
    
    %\item[\textbf{\resq{3}}] \textbf{[Plausibility] How plausible are generated change descriptions to domain experts?} We conduct a qualitative user study with domain experts in the field of the description model of \ac{sge}, to assess whether they approve our description approach and to which extent they see a practical application of our description approach. For that, we use a subset of our subject system and discuss exemplary change descriptions in a semi-structured interview.
    \item[\textbf{\resq{3}}] We conduct a qualitative user study with domain experts from the field of the description model of \ac{sge}, to assess three sub-\acp{rq}: \begin{itemize}[leftmargin=37pt]
        \item[\textbf{\resq{3.1}}] We evaluate \emph{plausibility} by answering the question: \textbf{How plausible are generated change descriptors, and the procedure to generate them, to participants?} We perform our description approach together with the participants along an exemplary subject system. In a semi-structured interview, we then assess whether the participants think the resulting change descriptors are convincing and whether they approve the procedure to generate them.
        \item[\textbf{\resq{3.2}}] We evaluate \emph{practical applicability} by answering the question: \textbf{To which extent do participants see practical application of our description approach?} We assess how supportive change descriptors are in model-based systems engineering and whether the participants see potential of our description approach to be used in automated system development processes.
        \item[\textbf{\resq{3.3}}] We evaluate \emph{extensibility} by answering the question: \textbf{Which possible extensions to our description approach are mentioned by participants?} We explicitly extract how change descriptors, and the procedure to generate them, can be extended to advance their plausibility and practical applicability, to lay foundations of our future work.
    \end{itemize}

    % OLD ---------
    %\item[\textbf{\resq{3}}] \textbf{[Plausibility] How plausible are generated change descriptions to domain experts?} We conduct a user study with domain experts, where we specify exemplary atomic and composed mappings, apply them to a single product evolution scenario, and discuss the resulting change descriptors and assessments in a semi-structured interview. 
    %We conduct a case study where change operations of a given product evolution scenario are translated into the description model of \ac{sge}. From experts (i.e., our participants who work within this domain), we elicit in interviews whether they approve our approach in context of the description model of \ac{sge}, and furthermore the extent to which they see the practical application of our approach, for example, whether as support or even as a substitute of manual effort for describing changes.
    
\end{itemize}
This section is organised as follows: We present the subject system we use for both our quantitative and qualitative evaluation in \autoref{sec:eval:subjectsystem}, before we present the actual evaluation of our three research questions; \resq{1} in \autoref{sec:eval:rq1}, \resq{2} in \autoref{sec:eval:rq2}, and \resq{3} in \autoref{sec:eval:rq3}. Lastly, we cover possible threats to validity in \autoref{sec:eval:threats}.

\input{01_chapters/evaluation_subject_system}

\input{01_chapters/evaluation_rq1}
\input{01_chapters/evaluation_rq2}
\input{01_chapters/evaluation_rq3}
\input{01_chapters/evaluation_threats}

%% file: 01_chapters/evaluation_subject_system.tex
\subsection{Subject System}
\label{sec:eval:subjectsystem}

As subject system for both research questions \resq{2} and \resq{3}, we use an adaption of the \ac{bcs} case study \cite{lityDeltaorientedSoftwareProduct2012}, a delta-oriented cyber-physical product line from the automotive industry. Expressing the user-experienceable functionality of a car, it specifies features of power windows, electric mirrors and an alarm system. From its modelled configurability with over 11000 different possible product variants, we use a representative subset of 18 product variants, pre-defined by Lity et al. \cite{lityDeltaorientedSoftwareProduct2012} along pairwise feature interaction coverage \cite{osterPairwiseFeatureinteractionTesting2011}. As concrete modelling artefacts, the case study provides specifications for \acp{cam} and \acp{smm}; precisely a core \ac{cam} and 25 \ac{cam} deltas, and a core \ac{smm} and 42 \ac{smm} deltas.

\begin{figure}
    \centering
    %\includesvg[width=1.0\linewidth]{02_figures/bcs_derivation_tree.svg}
    \includegraphics[width=1.0\linewidth]{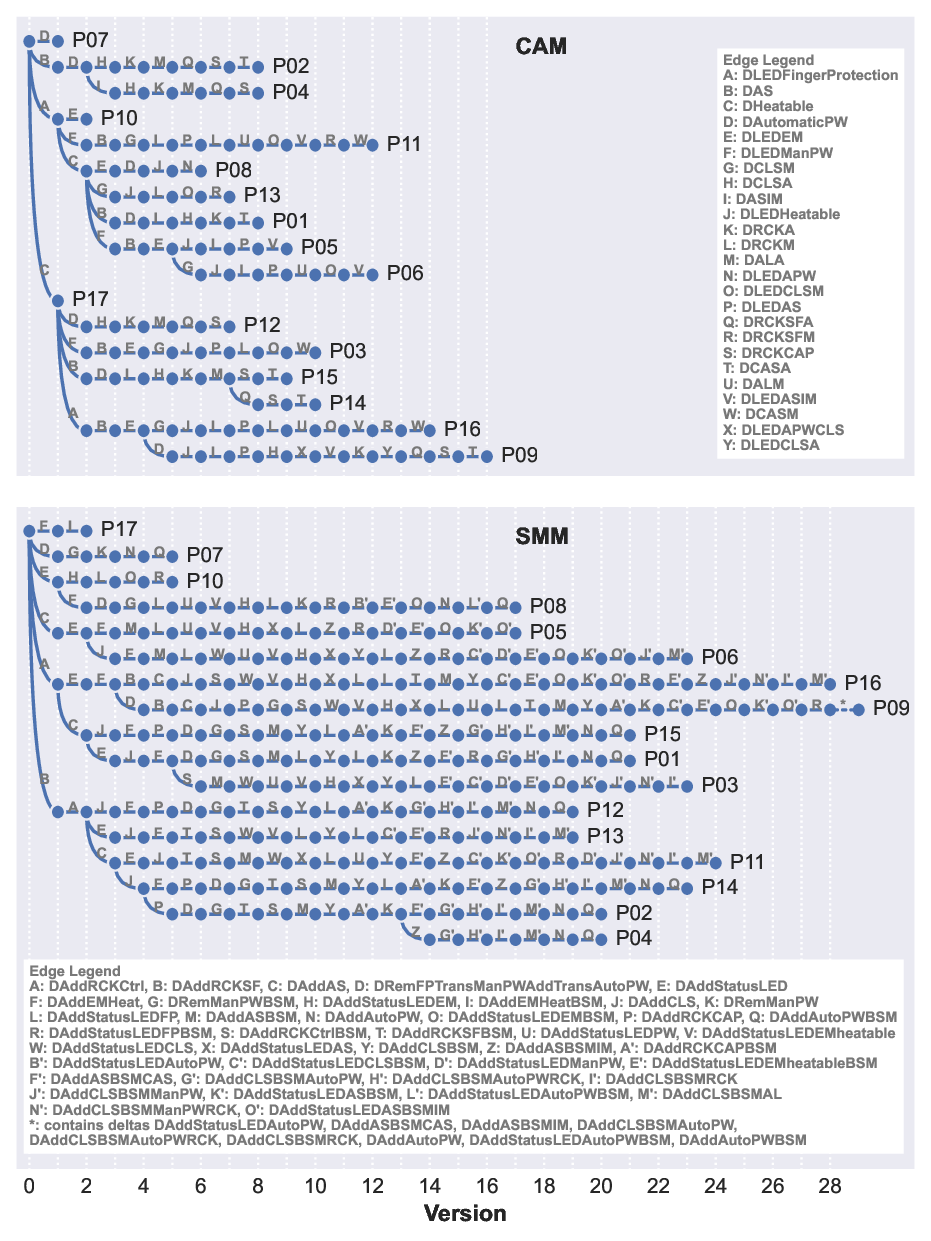}
    \caption{Evolution scenarios for \acp{cam} and \acp{smm} of 17 products, adapted from the \ac{bcs} case study \cite{lityDeltaorientedSoftwareProduct2012}.}
    \label{fig:bcs_derivation_tree}
\end{figure}

In the time-variable context of this paper, we apply the given deltas as version deltas so that, based on the core product variant \product{0} as common initial version \version{0}, 17 different product evolution scenarios are specified by individual sequences of applied deltas. For example, the \ac{cam} of product variant \product{14} then consists of ten applied deltas, representing its product evolution from version \version{0} to version \version{10}. \autoref{fig:bcs_derivation_tree} shows the evolution scenarios of \acp{cam} and \acp{smm} of all 17 products. In general, this results in a total of 465 applied deltas, thereof 140 applied deltas for \acp{cam} and 325 applied deltas for \acp{smm}. The applied deltas contain a total of 16092 applied delta operations, thereof 11575 applied delta operations for \acp{cam} and 4517 applied delta operations for \acp{smm}. All artefacts used are provided in our Zenodo package \cite{ochs_inter-disciplinary-description--changes--model-based-engineering_2026}.

%% file: 01_chapters/evaluation_rq1.tex
%\subsection{\resq{1}: Implementation in the Java Programming Language}
\subsection{\resq{1}: Technical Feasibility}
\label{sec:eval:rq1}

We first present the resulting implementation of our description approach, and then discuss design decisions we made in context of the technical feasibility of our approach.

\subsubsection{Result: The \emph{\cpscommunicator{}} Framework}

We implemented a generic framework called \cpscommunicator{}\footnote{\cpscommunicator{} on GitHub: \cpscomm}, which provides functionality to translate model-specific changes to model-independent change descriptors. The framework can be extended and used for multiple models, such as \acp{cam} and \acp{smm}, and model-independent languages, such as the description model of \ac{sge}. The interaction is based on a command line interface, whose set of commands can be individually extended as more models and model-independent languages are added. For example, a developer can specify a command to trigger the translation of certain model changes to a certain model-independent language.

\begin{figure}
    \centering
    %\includesvg[width=1.0\linewidth]{02_figures/cps_communicator_architecture.svg}
    \includegraphics[width=1.0\linewidth]{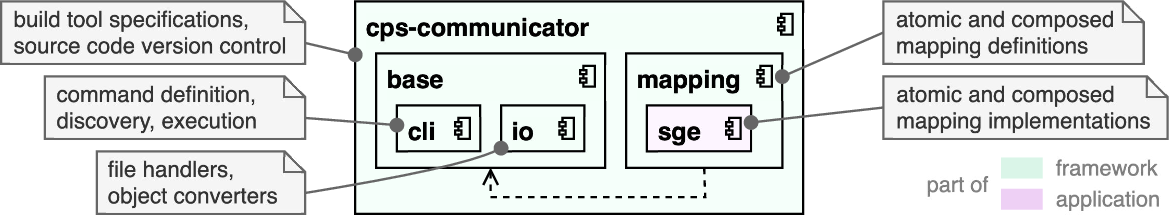}
    \caption{Component architecture and purposes of our \cpscommunicator{} framework (in green) and its applications (in purple).}
    \label{fig:cps_communicator_architecture}
\end{figure}

The \cpscommunicator{} framework consists of three main components, as illustrated in \autoref{fig:cps_communicator_architecture} in green: (1) The component \texttt{cps-communicator-base} defines the functionality of a command (interface \texttt{Command}, class \texttt{AbstractCommand}) and implements its discovery and execution, so that other components do not have to register individually. Further, the component implements basic utilities such as file handlers (class \texttt{JsonIO}) and object converters (class \texttt{EObjectIO}). (2) The component \texttt{cps-communicator-mapping} defines both atomic mappings (class \texttt{AtomicMapping}) and composed mappings (class \texttt{ComposedMapping}). Based on this, a developer can add concrete mappings between model-specific changes and model-specific languages, and with own commands to trigger them. We evaluate these steps in detail in our second research question on technical applicability, in \autoref{sec:eval:rq2}. (3) The third component \texttt{cps-communicator} provides technical infrastructure for managing all contained components, defines the entry point of the command line interface, and glues the framework together.

Technically, the \cpscommunicator{} framework is implemented in the Java programming language\footnote{Java Version \texttt{17}; \java}, so that it can be used on several execution platforms. Gradle\footnote{Gradle Version \texttt{8.10}; \gradle} is used as a build tool for managing the source code of the underlying components. A Gradle command \texttt{shadowJar} is specified which collects the components source code into a single Java archive (JAR-file), which itself can then be executed with Java in the command line. For source code version control, each component has its own Git\footnote{Git Version \texttt{2.42}; \git} repository, so that it can evolve independently from other components. The framework uses each component's version specified in the file \texttt{.gitmodules}, which itself is tracked by the Git repository of the component \texttt{cps-communicator}, enabling fine-grained version control over included and excluded components.

%Within the command line interface, a command specifically is an implementation of the abstract class \texttt{AbstractCommand} (component \texttt{cps-communicator-base}) and hence has a name, a (possibly empty) set of options, and an implementation of what it actually does when executed. The framework discovers and collects all implemented commands and displays them to a user when started. Furthermore, it can display all available options to a specific command.

Regarding modelling elements, i.e entities of a model, the \cpscommunicator{} framework relies on the definition from the interface \texttt{EObject} within the \ac{emf}\footnote{\ac{emf} Version \texttt{2.35}; \emf}. It provides necessary properties, such as object containment references.

Within the mapping implementations, an atomic mapping (class \texttt{AtomicMapping}) consists of a class reference representing a delta operation, and a reference to a generic change descriptor. In the application, a concrete instance of a delta operation is mapped to a concrete instance of a change descriptor. A composed mapping (class \texttt{ComposedMapping}) covers the application of atomic mappings to compute change descriptors with the method \texttt{getChangeDescriptors}, and the computation of an aggregated change descriptor with the method \texttt{getAggregatedChangeDescriptor}. For both of these methods, the composed mapping takes a set of delta operations, and, for the aggregated change descriptor additionally, the set of modelling elements of the updated model.

\subsubsection{Discussion: Design Decisions in Context of Technical Feasibility}

The aim behind our implementation, the \cpscommunicator{} framework, is to show general applicability of the underlying approach to arbitrary models and delta dialects, such as of \acp{cam} and \acp{smm}, and model-independent languages, such as the description model of \ac{sge}. Hence, we abstract from (1) models, (2) delta dialects, and (3) both atomic and composed mappings as far as possible, as discussed in the following parts. (1) We abstract from concrete models using a meta-model, because our concepts of composed mappings and atomic mappings are based on (changed) modelling elements within these models. For example, modelling elements of a \ac{cam} include components and connectors, while modelling elements of a \ac{smm} include states and transitions. We choose \ac{emf} as modelling element abstraction, because (a) with the interface \texttt{EObject}, it provides such an abstraction, with necessary identifiers and containment functionalities; and (b) because common modelling tool suites, such as Papyrus\footnote{\papyrus}, also build upon \ac{emf} with their \ac{uml} implementations. (2) We abstract from concrete delta dialects using a delta meta-model, because we require delta dialects to comply with general delta modelling concepts \cite{schaeferVariabilityModellingModelDriven2010}, such as delta operations derived from either additions, modifications or deletions. For example, modelling elements, such as components in a \ac{cam} or states in a \ac{smm}, can be added, modified and deleted. An abstract implementation of delta modelling is given in \emph{DeltaEcore}\footnote{\deltaecore} \cite{seidlDeltaEcoreAModelBasedDelta2014}, which itself is based on \ac{emf}, and used in our implementation for two reasons: (a) We can generate a concrete delta dialect automatically from a given meta-model within \ac{emf}, so by the ease of use we do not have to create it manually. (b) We can generate an according \emph{delta language} for a concrete delta dialect, which is an implementation of the delta dialect in a programming language for programmatic use, e.g. instantiable objects, in our case in the Java programming language. (3) We generalised atomic and composed mappings using a generic representation, because concrete mapping specifications have structure and functionality in common: For each atomic mapping, the structure is given by the assignment to one specific delta operation of a delta dialect, and to one specific change descriptor of a model-independent language; and the functionality is given by the application to a concrete delta operation contained in a delta. For each composed mapping in turn, the functionality is given by the application to a set of concrete delta operations. Regarding the results of the applied mappings, atomic mappings are a canonical set of change descriptors, and composed mappings are aggregated change descriptors. The latter is more difficult to represent generically, because it can represent a metric, such as the share of changed modelling elements, or even further sets of elements, for example all changed modelling elements itself. This means, while atomic mappings work solely on delta operations, a composed mapping may also take the set of modelling elements of the updated model into account, for example to compute the share of changed modelling elements. Thus, a composed mapping takes a set of delta operations \emph{and} a set of modelling elements of the updated model into account.

In summary, we answer \resq{1} as follows: We assume our approach as technically feasible; specifically, we achieved a full separation between abstract definitions and functionalities, given in the \cpscommunicator{} framework, and concrete implementations for application, further assessed in \autoref{sec:eval:rq2}.
%Besides this separation, the package structure allows us to also separate between the concerns of core functionality, command line interface, and technical management. Due to the nature of modelling concepts, we suggest using an object-oriented programming language, such as we did with the Java programming language, with decent abstraction capabilities. For example, the native usage of \ac{emf} in the Java programming language is an advantage over other programming languages which are either not object-oriented or do not have possibilities to handle \ac{emf} model data directly. However, with comprehensive parsing, we assume our approach to be technically feasible in other programming languages too.

%% file: 01_chapters/evaluation_rq2.tex
%\subsection{\resq{2}: Application to the \ac{bcs} Case Study}
\subsection{\resq{2}: Technical Applicability}
\label{sec:eval:rq2}

We quantitatively assess how atomic and composed mapping specifications can be applied to existing product evolution scenarios from the \ac{bcs} case study \cite{lityDeltaorientedSoftwareProduct2012, ochs_inter-disciplinary-description--changes--model-based-engineering_2026}. For that, we use and extent our implementation from \resq{1} so that we can translate model-specific changes into the description model of \ac{sge}.

In this section, we first describe our setup and the application procedure itself, and then present the resulting change descriptors, and finally discuss overall technical applicability.
%We quantitatively evaluate the technical applicability of our description approach to the subject system, the \ac{bcs} case study, based on our implementation efforts from \resq{1}. Here, we first describe our setup and the application procedure itself. Then we present the resulting change descriptors and assessments and finally discuss overall technical applicability.

% mention Zenodo here?

\subsubsection{Setup}
Along the \ac{bcs} case study, we describe the setup of delta dialects, of a meta-model for the description model of \ac{sge}, and of specifications of atomic and composed mappings, in the following paragraphs.

\begin{figure*}
    \centering
    \begin{subfigure}[t]{0.46\textwidth}
        \centering
        %\includesvg[width=1\linewidth]{02_figures/cam_meta_model.svg}
        \includegraphics[width=1\linewidth]{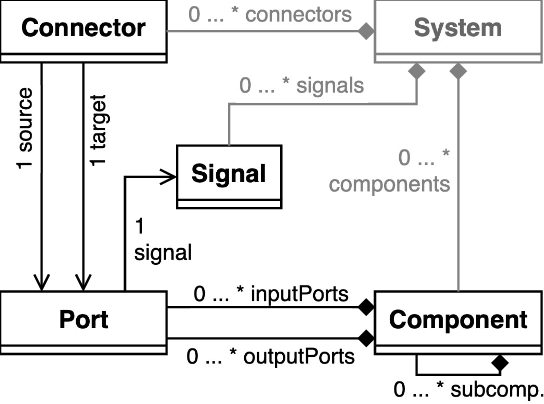}
        \subcaption{}
        \label{fig:cam_meta_model}
    \end{subfigure}%
    ~ 
    \begin{subfigure}[t]{0.54\textwidth}
        \centering
        %\includesvg[width=1\linewidth]{02_figures/smm_meta_model.svg}
        \includegraphics[width=1\linewidth]{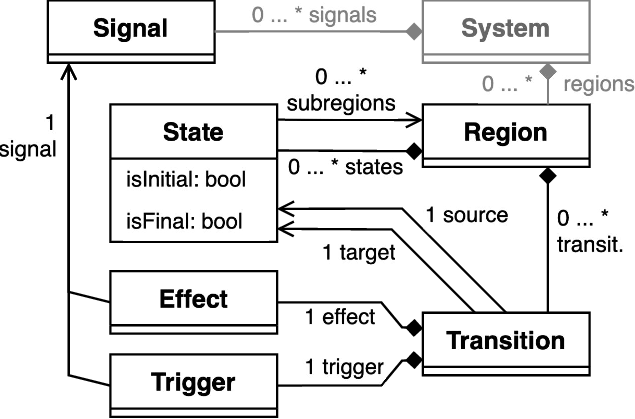}
        \subcaption{}
        \label{fig:smm_meta_model}
    \end{subfigure}
    \caption{Meta-model for \acp{cam} (a) and \acp{smm} (b).}
\end{figure*}

First, we make the underlying delta dialect for each model explicit, because it specifies all possible delta operations, and because our atomic mappings are specified upon these delta operations. However, instead of re-engineering each delta dialect from the \ac{bcs} case study by hand, we use a more forward-faced approach by \emph{deriving} the delta dialects from meta-models. More concretely, we first define \ac{emf}-based meta-models for (a) \acp{cam} and (b) \acp{smm}, such that we are able to express the instantiations given in the \ac{bcs} case study. For (a) \acp{cam}, illustrated in \autoref{fig:cam_meta_model}, we model entities for components, ports, connectors and signals, so that a component has input ports and output ports, and a connector connects one source port with one target port. A component can also contain subcomponents, and ports comply to a signal specification. For (b) \acp{smm}, illustrated in \autoref{fig:smm_meta_model}, we model entities for regions, transitions, states, triggers, effects and signals, so that regions can contain states, and states in turn can contain subregions. A state can be an initial state and/or a final state. A transition connects a source state with a target state. A transition can have triggers and effects, which comply to a signal specification. Next, we use \emph{DeltaEcore} \cite{seidlDeltaEcoreAModelBasedDelta2014} to derive the delta dialect for (a) \acp{cam} with 46 different delta operations, from which twelve are additions, 22 are modifications and twelve are deletions. For (b) \acp{smm} we derive a delta dialect with 50 different delta operations, from which ten are additions, 27 are modifications and 13 are deletions.

In a second step, we set up a meta-model of the model-independent description language to represent change descriptors for atomic mappings, and aggregated change descriptors for composed mappings. As we use the description model of \ac{sge}, we specifically model an entity for a variation which has a reference to a variation type, i.e. an \ac{av}, \ac{pv}, or \ac{cv}, for completeness.

Third, we specify atomic mappings to translate delta operations into the description model of \ac{sge}. For that we take a reduced delta dialect as a basis, as shown in \autoref{tab:no_delta_ops}: (a) From the \ac{cam} delta dialect, only 14 out of 46 defined delta operations were practically used in the \ac{bcs} case study: Five delta operations for addition, seven for modification, two for deletion of modelling elements. As the remaining 32 delta operations are never used in the \ac{bcs} case study, we do not consider them for our atomic mappings. Analogously, (b) from the \ac{smm} delta dialect, only 14 out of 50 defined delta operations were practically used in the \ac{bcs} case study: Eight delta operations for addition, five for modification, three for deletion of modelling elements. Generally, we aligned our atomic mapping specifications as described in our concept (\autoref{sec:concept:atomic_mappings}): A change of either a modelling element or a relation between two modelling elements is considered to be a \ac{pv}, while a change of an attribute of a modelling element is considered to be an \ac{av}. Thus, (a) for the delta operations of the reduced \ac{cam} delta dialect, we translate a delta operation into an \ac{av}, as long as it does not add, modify or delete either the source port or the target port of a connector. Else, we assume a change in the functional principle of the modelled system and hence translate the delta operation into a \ac{pv}. Analogously, (b) for the delta operations of the reduced \ac{smm} delta dialect, we translate a delta operation into an \ac{av}, as long as it does not add, modify or delete either the source state or target state of a transition; or does not add, modify or delete the trigger or effect of a transition; or does not modify the properties of a state being initial or final. Else, we assume a change in the behaviour of the modelled system and hence translate the delta operation into a \ac{pv}. \autoref{tab:atomic_mappings} shows our atomic mapping specifications for the delta operations contained in the reduced delta dialect of \acp{cam}. The 14 delta operations are mapped to ten \acp{av} and four \acp{pv}.

\begin{table*}[h]
\centering
\begin{tabular}{@{}l@{\hskip 1cm}l@{\hskip 1cm}llll@{}}
\toprule
\textbf{Model} & \multicolumn{5}{@{}l}{\textbf{no. Delta Operations}} \\
 &  & for add. & for mod. & for del. & $\sum$ \\ \midrule
\ac{cam} & in Delta Dialect & 12 & 22 & 12 & \textbf{46} \\
 & $\hookrightarrow$ in reduced Delta Dialect & $\hookrightarrow$ 5 & $\hookrightarrow$ 7 & $\hookrightarrow$ 2 & $\hookrightarrow$ \textbf{14} \\ \midrule
\ac{smm} & in Delta Dialect & 10 & 27 & 13 & \textbf{50} \\
 & $\hookrightarrow$ in reduced Delta Dialect & $\hookrightarrow$ 8 & $\hookrightarrow$ 5 & $\hookrightarrow$ 3 & $\hookrightarrow$ \textbf{14} \\ \bottomrule
\end{tabular}
\caption{Number of delta operations specified in delta dialect, and reduced delta dialect implied from the \ac{bcs} case study \cite{lityDeltaorientedSoftwareProduct2012}, for \acp{cam} and \ac{smm}.}
\label{tab:no_delta_ops}
\end{table*}

\begin{table*}[h]
\centering
\begin{subtable}{.65\linewidth}
% \raggedright % Left-align content
    \begin{tabular}{@{}l@{\hskip 3cm}cc@{}}
        \toprule
         & \multicolumn{2}{@{}c@{}}{\textbf{Variation Type}} \\
        \textbf{Delta Operation} & \ac{av} & \ac{pv} \\ \midrule
        \texttt{AddToComponentsInSystem} & x &  \\
        \texttt{AddToConnectorsInSystem} &  & x \\
        \texttt{AddToInputPortsInComponent} & x &  \\
        \texttt{AddToOutputPortsInComponent} & x &  \\
        \texttt{AddToSignalsInSystem} & x &  \\ \midrule
        \texttt{ModifyAttributeNameInComponent} & x &  \\
        \texttt{ModifyAttributeNameInConnector} & x &  \\
        \texttt{ModifyAttributeNameInPort} & x &  \\
        \texttt{ModifyAttributeNameInSignal} & x &  \\ 
        \texttt{SetSignalInPort} & x &  \\
        \texttt{SetSourcePortInConnector} &  & x \\
        \texttt{SetTargetPortInConnector} &  & x \\ \midrule
        \texttt{RemoveFromComponentsInSystem} & x &  \\
        \texttt{RemoveFromConnectorsInSystem} &  & x \\ \midrule
        $\sum$ & \textbf{10} & \textbf{4} \\ \bottomrule
    \end{tabular}
    \centering
    \caption{}
    \label{tab:atomic_mappings}
\end{subtable}%
\hfill
\begin{subtable}{.30\linewidth}
  % \raggedleft % Right-align content
    \begin{tabular}{@{}r@{\hskip 3cm}r@{}}
        \toprule
        \multicolumn{2}{c@{}}{\textbf{no. Applications}} \\
        AVs & PVs \\ \midrule
        134 &  \\
         & 1005 \\
        993 &  \\
        761 &  \\
        763 &  \\ \midrule
        134 &  \\
        1005 &  \\
        1754 &  \\
        763 &  \\
        2010 &  \\
         & 1005 \\
         & 1005 \\ \midrule
        21 &  \\
         & 222 \\ \midrule
        \textbf{8338} & \textbf{3237} \\ \bottomrule
    \end{tabular}
    \centering
    \caption{}
    \label{tab:atomic_mapping_applications}
\end{subtable}
\caption{14 delta operations specified in the reduced delta dialect for \acp{cam} with (a) their atomic mapping specifications and (b) their number of applications in the \ac{bcs} case study.}
\end{table*}

In a fourth step, we specify composed mappings, to generate aggregated change descriptors for each delta. Specifically, we take the idea of assessing the development risk incorporated in a delta, described in our concept (\autoref{sec:concept:composed_mappings}), by expressing (a) the extent to which a delta is changing a model, and (b) the proportion of variation types which describe the contained delta operations within the delta.

To technically execute our setup, we use our implementation from \resq{1}, the \cpscommunicator{} framework. We extent the framework with the component \texttt{cps-communicator-mapping-sge}, as illustrated in purple in \autoref{fig:cps_communicator_architecture}. Within this new component, we implement the class \texttt{SGEComposedMapping}, which realises the computation of the share of changed modelling elements and the proportion of variation types. We further implement commands for the command line interface of the framework to trigger the application of specified atomic and composed mappings to the \acp{cam} (\texttt{ApplyDeltasComponentCommand}) and \acp{smm} (\texttt{ApplyDeltasStatemachineCommand}) of the \ac{bcs} case study.

\subsubsection{Procedure}
For \acp{cam} and \acp{smm} of the \ac{bcs} case study, we iterate over all 17 products per model, and over all applied deltas per product, inducing the product's evolution. For each delta, we apply the composed mapping, to get the share of changed modelling elements and the proportion of variation types. For each delta operation contained in a delta, we apply the according atomic mapping, to get the variation type within the description model of \ac{sge}.

\subsubsection{Results}
Based on our procedure, the results, provided in our Zenodo package \cite{ochs_inter-disciplinary-description--changes--model-based-engineering_2026}, are structured along \acp{cam} and \acp{smm}. Specifically, we present data of the applied deltas (140 for \acp{cam}, 325 for \acp{smm}) and their contained delta operations (11575 for \acp{cam}, 4517 for \acp{smm}) with regard to the share of changed modelling elements and the proportion of variation types.

\begin{figure}
    \centering
    %\includesvg[width=1.0\linewidth]{02_figures/scatter_deltas_to_no_me_and_socme.svg}
    \includegraphics[width=1.0\linewidth]{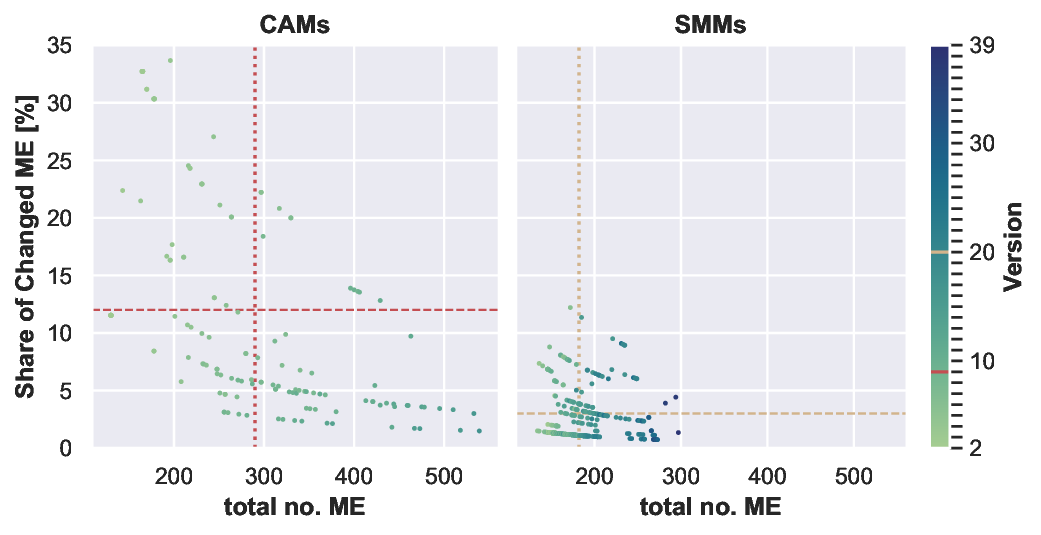}
    \caption{Share of changed modelling elements in relation to total number of modelling elements for each applied delta in \acp{cam} and \acp{smm}.}
    \label{fig:scatter_deltas_to_no_me_and_socme}
\end{figure}

In \autoref{fig:scatter_deltas_to_no_me_and_socme}, we scattered applied deltas to the the number of modelling elements related to the share of changed modelling elements. The left scatter plot belongs to \acp{cam}, the right to \acp{smm}. The horizontal axes represent the number of modelling elements after a delta was applied, and have the same range for \acp{cam} and \acp{smm}, to make absolute model sizes comparable. The vertical axis represents the share of changed modelling elements in percent, caused by a delta application. The hue of a delta represents the version in which the delta is applied to a model, while greenish colours represent earlier versions and blueish colours represent later versions. Both scatter plots show a trend towards an increased number of modelling elements in later versions. With an average number of modelling elements of 290 (red dotted line), \acp{cam} have a higher number of modelling elements than \acp{smm}, with an average number of modelling elements of 182 (beige dotted line). Furthermore, with an average share of changed modelling elements of $12\; \%$ (red dashed line), \ac{cam} deltas cause a higher share of changed modelling elements than \ac{smm} deltas, with an average share of changed modelling elements of $3\; \%$ (beige dashed line). In turn, with an average number of versions of 20 (beige solid line in colour bar), \acp{smm} have a higher number of versions than \acp{cam}, with an average number of versions of 9 (red solid line in colour bar). Overall, both scatter plots show a trend for a decreasing share of changed modelling elements for later versions and for higher numbers of modelling elements.

%Regarding \acp{cam}, there are in total 140 deltas and in total 11575 delta operations. In an average delta, $\approx 28$ modelling elements out of $\approx 290$ are changed, with a resulting share of changed modelling elements of $\approx 12\; \%$. It furthermore consists of $\approx 83$ delta operations, from which $\approx 53$ ($\approx 64\; \%$) are mapped to \acp{av} and $\approx 30$ ($\approx 36\; \%$) are mapped to \acp{pv}.

%Regarding \acp{smm} of the 17 investigated product evolution scenarios, there are in total 325 deltas and in total 4517 delta operations. In an average delta, $\approx 5$ modelling elements out of $\approx 182$ are changed, with a resulting share of changed modelling elements of $\approx 3\; \%$. It furthermore consists of $\approx 14$ delta operations, from which $\approx 8$ ($\approx 57\; \%$) are mapped to \acp{av} and $\approx 6$ ($\approx 43\; \%$) are mapped to \acp{pv}.

The proportion of variation types across all applied delta operations was in average $72 \; \%$ \acp{av} to $28 \; \%$ \acp{pv} for \acp{cam}, and $62 \; \%$ \acp{av} to $38 \; \%$ \acp{pv} for \acp{smm}. For applied delta operations in \acp{cam} in detail, \autoref{tab:atomic_mapping_applications} shows how often each delta operation of the reduced \ac{cam} delta dialect was applied and thus mapped to either an \ac{av} (8338 times, $\approx 72 \; \%$) or a \ac{pv} (3237 times, $\approx 28 \; \%$).

For product \product{14} as a concrete example of an evolution scenario, \autoref{tab:ex_evolution_scenario} shows collected data of its ten applied deltas from version \version{0} to version \version{10}. The ten applied deltas consist of a total of 730 contained delta operations. The number of contained delta operations per applied delta decreases almost monotonously across the versions, from 175 (delta \texttt{DHeatable}, \versionjump{1}{2}) to 20 (delta \texttt{DCASA}, \versionjump{10}{11}), with a two exceptions (delta \texttt{DCLSA}, \versionjump{5}{6}, and delta \texttt{DRCKCAP}, \versionjump{9}{10}). With atomic mappings, the 730 delta operations are mapped to a total of 518 \acp{av} and 212 \acp{pv}, resulting in a proportion of \acp{av} to \acp{pv} of $71 \; \%$ to $29 \; \%$. The total number of modelling elements is increasing monotonously, from 165 in version \version{1} (after application of delta \texttt{DHeatable}) to 342 in version \version{10} (after application of delta \texttt{DCASA}), while the number of changed modelling elements decreases, again almost monotonously, from 54 in version \version{1} (after application of delta \texttt{DHeatable}) to 8 in version \version{10} (after application of delta \texttt{DCASA}). This results in an also decreasing share of changed modelling elements from $33 \; \%$ in version \version{1} to $2 \; \%$ in version \version{10}.

\begin{table*}[h]
\centering
\newcommand*\rot{\multicolumn{1}{A{35}{0em}}}
\begin{tabular}{@{}l@{\hskip 0.15cm}lr@{\hskip 0.8cm}r@{\hskip 0.1cm}rr@{\hskip 0.1cm}r@{\hskip 0.6cm}rr@{\hskip 0.75cm}r@{}}
\toprule
 &  &  & \multicolumn{4}{l}{\textbf{no. Variation Types}} & \multicolumn{3}{l}{\textbf{no. Model. Elem.}} \\
 & \textbf{Delta} & \rot{\textbf{no. DOp*}} & \multicolumn{2}{c}{\textbf{AVs}} & \multicolumn{2}{c}{\textbf{PVs}} & \rot{\textbf{total}} & \rot{\textbf{changed}} & \textbf{SoCME**} \\ \midrule
\versionjump{0}{1} & \small{DHeatable} & 175 & 115 & ($66 \%$) & 60 & ($34 \%$) & 165 & 54 & $\approx\!33 \%$ \\
\versionjump{1}{2} & \small{DAS} & 158 & 119 & ($75 \%$) & 39 & ($25 \%$) & 218 & 53 & $\approx\!24 \%$ \\
\versionjump{2}{3} & \small{DAuto.PW} & 99 & 64 & ($65 \%$) & 35 & ($35 \%$) & 245 & 32 & $\approx\!13 \%$ \\
\versionjump{3}{4} & \small{DASIM} & 36 & 27 & ($75 \%$) & 9 & ($25 \%$) & 257 & 12 & $\approx\!5 \%$ \\
\versionjump{4}{5} & \small{DCLSA} & 70 & 52 & ($74 \%$) & 18 & ($26 \%$) & 280 & 23 & $\approx\!8 \%$ \\
\versionjump{5}{6} & \small{DRCKA} & 50 & 38 & ($76 \%$) & 12 & ($24 \%$) & 297 & 17 & $\approx\!6 \%$ \\
\versionjump{6}{7} & \small{DALA} & 48 & 36 & ($75 \%$) & 12 & ($25 \%$) & 313 & 16 & $\approx\!5 \%$ \\
\versionjump{7}{8} & \small{DRCKSFA} & 24 & 18 & ($75 \%$) & 6 & ($25 \%$) & 321 & 8 & $\approx\!2 \%$ \\
\versionjump{8}{9} & \small{DRCKCAP} & 50 & 35 & ($70 \%$) & 15 & ($30 \%$) & 336 & 16 & $\approx\!5 \%$ \\
\versionjump{9}{10} & \small{DCASA} & 20 & 14 & ($70 \%$) & 6 & ($30 \%$) & 342 & 8 & $\approx\!2 \%$ \\ \midrule
$\sum$ & & 730 & 518 & ($71 \%$) & 212 & ($29 \%$) &  &  & \\ \bottomrule
\multicolumn{10}{r}{\footnotesize{* DOp: Delta Operations; ** SoCME: Share of Changed Modelling Elements}} \\\bottomrule
\end{tabular}
\caption{\ac{cam} evolution scenario for product \product{14} with 10 applied deltas and 730 contained delta operations.}
\label{tab:ex_evolution_scenario}
\end{table*}

\subsubsection{Discussion}

The technical application of our approach to the subject system led to the expected results, i.e. all 16092 delta operations were translated into change descriptors by atomic mappings, and all 465 deltas were translated into aggregated change descriptors by composed mappings.

The share of changed modelling elements typically decreases over time, as shown by the scatter plots in \autoref{fig:scatter_deltas_to_no_me_and_socme} and shown by the exemplary evolution of product \product{14} in \autoref{tab:ex_evolution_scenario}. We account that trend to deltas containing mostly additive delta operations instead of subtractive, shown for \acp{cam} in \autoref{tab:atomic_mapping_applications}. This means, the models get larger in terms of their number of modelling elements, but the applied deltas do not change proportionally as many of the modelling elements as the models grow.

Regarding the proportion of variation types, \acp{av} typically occur more often, in average for $72\;\%$ in \acp{cam} and for $62\;\%$ in \acp{smm}. This is a consequence of our atomic mapping specifications, where we use \acp{pv} to describe changes in a system's functionality (\ac{cam}) or behaviour (\ac{smm}). The resulting, lower proportion of \acp{pv} in the application then means, that roughly two thirds of the changes mainly concern the models attributes (\ac{av}), and not their function principles (\ac{pv}). For example, regarding \autoref{tab:atomic_mappings} and \autoref{tab:atomic_mapping_applications}, four out of 14 specified delta operations of the reduced \ac{cam} delta dialect are mapped to a \acp{pv}, and are applied 3237 out of 11575 times in total.

In summary, we answer \resq{2} as follows: Our approach is technically applicable; specifically, by extending our implementation, the \cpscommunicator{} framework, we enabled the shift, from manually translating 16092 applied delta operations of our subject system, into specifying 28 atomic mappings once, and then apply them 16092 times to the occurring delta operations.

%% file: 01_chapters/evaluation_rq3.tex
\subsection{\resq{3}: Plausibility, Practical Applicability \& Extensibility}
\label{sec:eval:rq3}

To assess \resq{3}, we conduct a qualitative user study, where participants apply our description approach to a product evolution scenario, and then are asked to answer questions in semi-structured interviews. Specifically, our subject system is a product evolution scenario of a \ac{cam}, taken from the \ac{bcs} case study. The model-independent language used is the description model of \ac{sge}, thus our participants are directly from this research field. We first let the participants specify atomic mappings for a subset of the \ac{cam} delta dialect, into the description model of \ac{sge}, and then apply their mapping specifications to the product evolution scenario. In the interviews, we obtain the participants opinions, to extract findings, and to finally discuss our three sub-\resq{}s \resq{3.1}, on plausibility, \resq{3.2}, on practical applicability, and \resq{3.3}, on extensibility.

\subsubsection{Participants}
\label{sec:eval:rq3:participants}

We worked together with twelve participants directly from the research field of the description model of \ac{sge}, because we used the description model of \ac{sge} as model-independent language in this study. Furthermore, within this research field, we assumed that the participants have an elementary understanding of artefacts from model-based systems engineering, which the \ac{bcs} case study formally represents.
%As the results of our mapping approach are change descriptors and change assessments within the description model of \ac{sge}, namely different variation types and the share of changed modelling elements, we worked together with twelve participants directly from the research field of the description model of \ac{sge}. Within this research field, we furthermore assumed that the participants have an elementary understanding of artefacts from model-based systems engineering, which the \ac{bcs} case study formally represents.

From each participant, we collected demographic information a small, unmoderated questionnaire upfront, to get a more fine-grained insight into our participants background, in relation to the description model of \ac{sge} and model-based systems engineering. Specifically, we asked about each participant's \emph{experience with the description model of \ac{sge}} in years (intervals; \category{less than 1 year}, \category{1 to 3 years}, \category{3 to 10 years}, \category{more than 10 years}), and for \emph{self-assessed knowledge in the description model of \ac{sge}}, in the Dreyfuss Skill Acquisition scale \cite{dreyfusFiveStageModelMental1980} (ordinal; \category{Novice}, \category{Advanced Beginner}, \category{Competent}, \category{Proficient}, \category{Master}). Second, we asked about each participant's \emph{self-assessed knowledge in model-based systems engineering}, again in in the Dreyfuss Skill Acquisition scale \cite{dreyfusFiveStageModelMental1980}, to see how strong experience and knowledge in the description model of \ac{sge} correlates to knowledge in model-based systems engineering. Third, we asked about each participant's \emph{working relationship}, were we distinguished between academia (categorial; \category{Student}, \category{(Post-) Doctoral Researcher}, \category{Professor}) and industry (\category{Engineer}, \category{Manager}).

\begin{figure}
    \centering
    %\includesvg[width=1.0\linewidth]{02_figures/demographic_overview.svg}
    \includegraphics[width=1.0\linewidth]{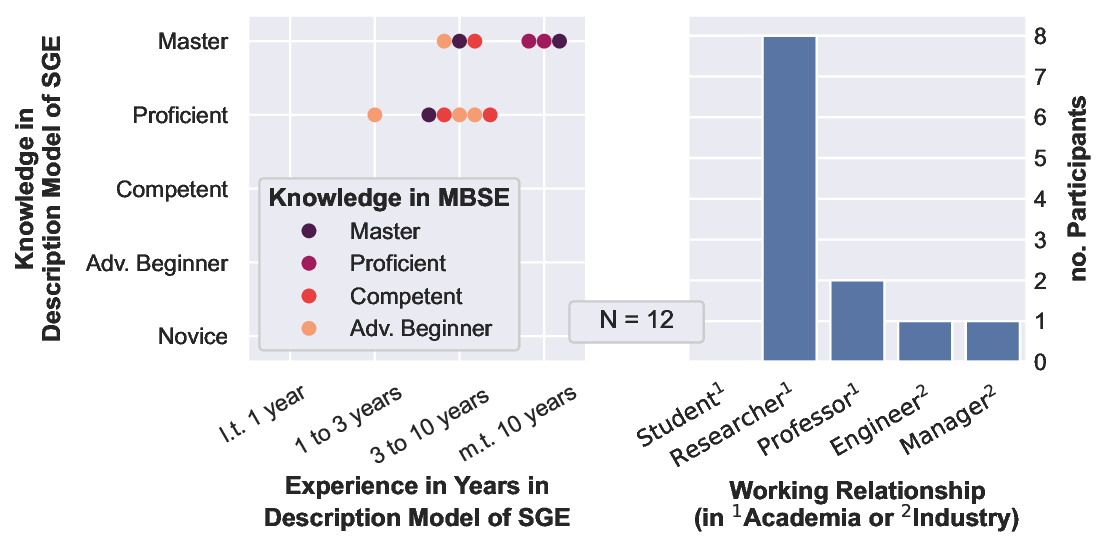}
    \caption{Overview of demographic information about participants.}
    \label{fig:demographic_overview}
\end{figure}

\autoref{fig:demographic_overview} shows the results of the demographic questionnaire. On the left, we scattered each participant's experience with the description model of \ac{sge} (horizontal axis) to the participant's knowledge in the description model of \ac{sge} (vertical axis). The hue represents the participant's knowledge in model-based systems engineering. We observe that all three dimensions correlate positively, meaning the more experience participants have with the description model of \ac{sge}, the more knowledge in the description model of \ac{sge} (Pearson correlation 0.603) and the more knowledge in model-based systems engineering (Pearson correlation 0.554) are they assessing themselves. On the right in \autoref{fig:demographic_overview}, the number of participants in each working relationship category are shown. From academia participated zero students, eight (post-) doctoral researchers and two professors. From industry participated one engineer and one manager.

\subsubsection{Subject System}
\label{sec:eval:rq3:subjectsystem}
As a subject system for our user study, we used a subset of the \ac{bcs} case study; in particular, the evolution of the \ac{cam} of product \product{14}, from version \version{0} to version \version{2}. This is because, first, we assumed that three versions (\version{0}, \version{1}, \version{2}) suffice for the purpose of showing our approach without mentally overloading the participants. Second, we assumed that \acp{cam} are known to the participants, as mechanical/ system engineers, better than \acp{smm}, so they should familiarise faster with the setting of the evaluation. Third, we used product \product{14}, because we assume the first two applied \ac{cam} deltas of product \product{14}, delta \texttt{DHeatable} and delta \texttt{DAS} (see \autoref{tab:ex_evolution_scenario}), to be illustrative for an exemplary model evolution. Specifically, delta \texttt{DHeatable} exchanges a component for a non-heatable electric mirror with a component for a heatable electric mirror, and delta \texttt{DAS} add a component for an alarm system.
%The second reason for choosing product \product{14} is, because delta \texttt{DHeatable} superficially modifies the existing non-heatable electric mirror component to a heatable one, but technically removes the non-heatable electric mirror component and completely adds a new heatable one. This results in a much higher number of changed modelling elements compared to a sole modification of the electric mirror component. We assume that participants 

As for the specification of atomic mappings, we further tried to reduce the mental and temporal load of the participants by only let them specify six out of 14 atomic mappings for the reduced \ac{cam} delta dialect. We included two delta operations for additions, namely (1) adding an input port to a component and (2) adding a connector; two delta operations for modifications, namely (3) modifying the name of a component and (4) modifying the source port of a connector; and two delta operations for deletions, namely (5) removing a connector and (6) removing a component. We took the remaining eight atomic mappings from our setup of \resq{2}, where we translate a delta operation into an \ac{av}, as long as it does not add, modify, or delete either the source port, or the target port of a connector. Else, we assume a change in the functional principle of the modelled system, and hence translate the delta operation into a \ac{pv}.

In summary, we prepared six delta operations to be mapped by the participant, and complement them with eight delta operations mapped by ourselves. With these, in total 14 atomic mapping specifications, the deltas for exchanging a component for a non-heatable electrical mirror with a heatable one (delta \texttt{DHeatable}) and for adding an alarm system (delta \texttt{DAS}) can be described within the evolution of product \product{14} from version \version{0} to version \version{2}.

\begin{table*}[h]
\centering
\begin{tabular}{lccr}
\toprule
\multicolumn{2}{l}{\textbf{We ...}} & \multicolumn{2}{r}{\textbf{The Participant ...}} \\ \midrule
\multicolumn{4}{c}{\textit{\textbf{Introduction Phase}}} \\
\multicolumn{4}{c}{\small{Material: demographic questionnaire, visual explanation of research rationale}} \\ \midrule
\textbf{1} & \multicolumn{2}{l}{give an introduction on goals and course of the study.}\qquad &  \\
 & \multicolumn{2}{r}{is asked to about demographic information.} & \textbf{2} \\
\textbf{3} & \multicolumn{2}{l}{give an introduction on \acp{cam}.} &  \\[5pt] \midrule
\multicolumn{4}{c}{\textit{\textbf{Specification Phase}}} \\
\multicolumn{4}{c}{\small{Material: illustration of delta operations, confidence questionnaire}} \\ \midrule
\textbf{4} & \multicolumn{2}{l}{show six selected delta operations.} & \\
 & \multicolumn{2}{r}{is asked to specify atomic mappings.} & \textbf{5} \\
 & \multicolumn{2}{r}{is asked about confidence in specifying atomic mappings.} & \textbf{6} \\[5pt] \midrule
\multicolumn{4}{c}{\textit{\textbf{Application Phase}}} \\
\multicolumn{4}{c}{\small{Material: illustration of evolution scenario}} \\ \midrule
 & \multicolumn{2}{r}{is asked to take a short break.} & \textbf{7} \\
\textbf{8} & \multicolumn{2}{l}{digitalise atomic mapping spec., generate change descriptors.}\qquad & \\
\textbf{9} & \multicolumn{2}{l}{show and explain the evolution scenario.} & \\[5pt] \midrule
\multicolumn{4}{c}{\textit{\textbf{Interview Phase}}} \\
\multicolumn{4}{c}{\small{Material: none}} \\ \midrule
\textbf{10} & ask interview questions / & \hskip -0.3cm is asked to to answer interview questions. & \textbf{10} \\
%\multirow{2}{*}{\textbf{10}} & \multicolumn{2}{l}{ask interview questions.} & \multirow{2}{*}{\textbf{10}} \\
% & \multicolumn{2}{r}{is asked to to answer interview questions.} & \\
\textbf{11} & \multicolumn{2}{l}{close the session.} & \\ \bottomrule
\end{tabular}
\caption{Session procedure and material of our user study.}
\label{tab:session_procedure}
\end{table*}

\subsubsection{Session Procedure \& Material}
\label{sec:eval:rq3:session_procedure}
We conducted the study \emph{individually} with each participant, so that direct influences between participants are mitigated, as the participants perceived different knowledge and experience levels among them (see \autoref{fig:demographic_overview} on the left). Furthermore, with individual sessions, the existing causality from the specified atomic mappings to the generated change descriptors is personalised to the participant, which we assumed to provide better insights into the participant's understanding of the matter.

To further investigate how the participants were specifying atomic mappings, we explicitly asked them how confident they were thereby, directly after, in a questionnaire. We prelimited options to (ordinal) \category{never confident}, \category{sometimes confident}, \category{mostly confident}, and \category{always confident}.

The material of the sessions comprises of the demographic questionnaire, an optional visual explanation of the research rationale, an illustration of all delta operations to be translated into the description model of \ac{sge} with atomic mappings, a questionnaire about the participant's confidence in specifying atomic mappings, and an illustration of the evolution scenario.

We prepared the sessions to be done in person and online. For the sessions in person, we printed necessary material on paper, so that the participants could lay out and perceive the information as suitable. For the sessions online, we used Microsoft Teams\footnote{Microsoft Teams: \teams{}} for audio-visual communication, and Miro\footnote{Miro: \miro{}} as virtual desk.
%Nine out of twelve sessions were done in person and with the material printed on paper, so that the participants could lay out and perceive the information as suitable. Three sessions were done online, due to logistical reasons, with tool support of Microsoft Teams\footnote{Microsoft Teams: \teams{}} for audio-visual communication and Miro\footnote{Miro: \miro{}} as virtual desk.

%The material of the sessions comprised of the demographic questionnaire, the illustration of all delta operations to be translated into the description model of \ac{sge}, the question about the participant's confidence for ..., and the illustration of the three investigated versions of the subject system. In the sessions conducted online, we additionally give a visual explanation of the research rationale.

Each session's procedure was intended to take between 45 and 60 minutes, and was structured along an introduction, specification, application and interview phase, as shown in an overview in \autoref{tab:session_procedure}, and presented in detail in the following enumeration.

\begin{enumerate}
    \item[] \textit{Introduction Phase (short)}
    \item We give an introduction on the higher level goals and the course of the study, so that the participant can contextualise the work and understand the research rationale.
    \item The participant is asked to fill out the questionnaire on demographic information.
    \item We give an introduction on \acp{cam}, so that the participant can get familiar with the purpose of \acp{cam} and the elements within \acp{cam}, because the following tasks in the study are mainly based on \acp{cam}.\newline
    
    \item[] \textit{Specification Phase (extensive)}
    \item We show each of the six selected delta operations, one after another, and ...
    \item ... the participant is asked to specify the according atomic mapping, into a variation type (\ac{av} or \ac{pv}) within the description model of \ac{sge}.
%    \item We further ask for a reasoning behind the participant's decision, so that we can elicit further expert knowledge about the mapping specification.
    \item The participant is asked to fill out the questionnaire about confidence in specifying the atomic mappings.\newline
    
    \item[] \textit{Application Phase (short)}
    \item We ask the participant to take a short break while ...
    \item ... we digitalise the obtained atomic mapping specifications into our notebook, generate the change descriptors and aggregated change descriptors with the notebook, and wrote them back on paper for the evolution scenario.
    \item We show the evolution scenario to the participant and explain it orally in detail, so that the participant can understand on a higher level which changes the evolution scenario represents. We include the generated change descriptors and aggregated change descriptors in our explanations already, so that the participant can understand the relation between the change descriptors and aggregated change descriptors and our oral explanation of the evolution scenario.\newline

    \item[] \textit{Interview Phase (extensive)}
    \item We ask the participant to answer some questions about the passed study, i.e. give the interview, so that we can elicit desired information about the participant's opinions on the change descriptors and aggregated change descriptors, and our approach itself.
    \item We close the session.
\end{enumerate}

\noindent During the case study we answered questions advancing the participant's understanding, and gave support where appropriate, as long as it would not have directly led to answers for which we asked the participant originally.

\subsubsection{Interview Guide}
\label{sec:eval:rq3:interview_guide}
In the interview phase of our user study (step 10), we pose three questions to each participant, aimed to answer \resq{3}. Specifically, question Q10 refers to \resq{3.1} on plausibility, and is bound directly to the generated change descriptors of the treated evolution scenario, while Questions Q20 and Q21 refer to \resq{3.2} on practical applicability, and address our description approach more generally. \resq{3.3} on extensibility is covered implicitly by all interview questions. The interview questions are as follows:

\begin{itemize}[leftmargin=24pt]
    \item[\textbf{Q10}] How plausible are the change descriptors, and the procedure to generate them, in an \ac{sge} context, to you?
    \item[\textbf{Q20}] How supportive is the approach to describe changes within an model-based systems engineering context to you?
    \item[\textbf{Q21}] To what extent do you think the approach could replace manual effort within an model-based systems engineering context?
\end{itemize}

\noindent With Question Q10, we elicit whether the participants can relate the generated change descriptors to the concrete product evolution scenario, e.g. if a proportion of 25\% \acp{pv}, to exchange a non-heatable electric mirror with a heatable one, makes sense to them. Additionally, we elicit whether the participants understand and approve the procedure to generate the change descriptors, e.g. if atomic mapping specifications for delta operations into variation types makes sense to them. With Question Q20, we elicit to which extent and for which purposes the participants see an information gain in the generated change descriptors, in context of model-based systems engineering; for example, a descriptive purpose for documenting changes. With Question Q21, we elicit whether the participants see potential of our description approach to be applied in automated processes, e.g. automatically generate documentation without experts additions.

%Third, Question Q10 xxx us to assess whether the participants think that the procedure of our our approach makes sense or not, e.g. if the specification of atomic mappings for single delta operations is useful or too fine-grained. Then, Question Q20 helps us to assess whether the participants perceive an information gain in the generated change descriptors and assessments, in the use case of describing changes in model-based systems engineering; for example, a descriptive purpose for documenting changes. Finally, Question Q21 helps us to, first, assess whether the participants see application potential in our approach without manual, expert-based efforts, e.g. automatically generate documentation without experts additions. Second, Question Q21 helps us to assess whether the participants could imagine a practical application of our approach in automated processes, e.g. automatically making decisions in situations with several solution alternatives.

\subsubsection{Data Analysis}
\label{sec:eval:rq3:data_analysis}

%The transcripts are provided in the Hoffmann-Riem flavour \cite{hoffmann-riem_adoptierte_1998} in our replication package.
We recorded audio from all sessions and transcribed (audio-to-text) with \emph{aTrain}\footnote{aTrain Version \texttt{1.2.4}; \atrain{}}, a locally running, artificial-intelligence-driven transcription tool from academia, to comply with privacy regulations. We then analysed the transcripts with \emph{MAXQDA}\footnote{MAXQDA Version \texttt{24.11.1}; \maxqda{}} tool support.

Specifically, from the transcripts, we picked out segments in which participants gave answers to our interviews questions. We inductively extracted \emph{codes} and code categories from these segments, so that code categories refer to our interview questions, and codes within a code category represent different answers to the associated interview question. We denote code categories enumerated and with reference to the numbered interview question, e.g. \codecategory{10-0}, \codecategory{10-1} etc. refer to interview Question Q10. We denote codes enumerated and with reference to the belonging code category, e.g. \code{10-0}{A}, \code{10-0}{B} etc. refer to code category \codecategory{10-0}. For each code, we record the number of marked segments from transcripts absolute, and relative in relation to the whole code category the code belongs to (henceforth called \emph{evidence}). We further give the number of participants for which the code in question occurred. This way we can identify majorities among answers to our interview questions, per code category.

\subsubsection{Findings}
\label{sec:eval:rq3:findings}
We present the findings of our user study structured along the session procedure. Specifically, we cover the atomic mapping specifications of the specification phase, and the participants answers to the questions of the interview phase. In our Zenodo package \cite{ochs_inter-disciplinary-description--changes--model-based-engineering_2026}, we provide the collected atomic mapping specifications and generated change descriptors of all sessions, as well as transcripts in the Hoffmann-Riem flavour \cite{hoffmann-riemAdoptierteKindFamilienleben1998}, a code matrix of the interview answers and evidence for all findings presented in this section.

\paragraph{Specification Phase}

In total, ten out of twelve participants performed the atomic mapping specifications. Two participants did not, for split reasons: Either they expected a contextualisation of all presented delta operations into an application context already, by arguing about a missing reference system, or they would have mapped some delta operations to \acp{cv}, which was not intended by our study design. In both cases, we used our own atomic mapping specifications from \resq{2} for the application to the evolution scenario. In the application phase, both participants were then introduced to change descriptors based on our own atomic mapping specifications. This way, even if participants did not specify mappings before, we could show actual change descriptors for the presented evolution scenario to \emph{all} participants, and then start the interview phase on the same foundation, for every participant.

For the ten participants who did the atomic mapping specifications, \autoref{tab:mapping_scenario_results} shows the resulting distribution and standard deviation\footnote{\label{footnote:standard_deviation}standard deviation ($\sigma$) calculated by encoding \ac{av} as 0, and \ac{pv} as 1} among the participants, for each of the six delta operations mapped to either \ac{av} or \ac{pv}. Delta operation (3), for modifying the name of a component, was mapped unanimously to an \ac{av} by all participants. Delta operations (4), for modifying the source port of a connector, and (5), for removing a connector, were mostly (standard deviation $0.4$) mapped to \acp{pv}. Delta operations (2), for adding a connector, and (6), for removing a component, show increased scattering (standard deviation $0.458$), mapped to \ac{av} and \ac{pv}. Delta operation (1), for adding an input port to a component, was equally often mapped to an \ac{av} as well as a \ac{pv} (standard deviation $0.5$), showing the highest scattering among the participants.

\begin{figure}[h]
    \centering
    %\includesvg[width=0.5\linewidth]{02_figures/hist_mapping_confidence.svg}
    \includegraphics[width=0.5\linewidth]{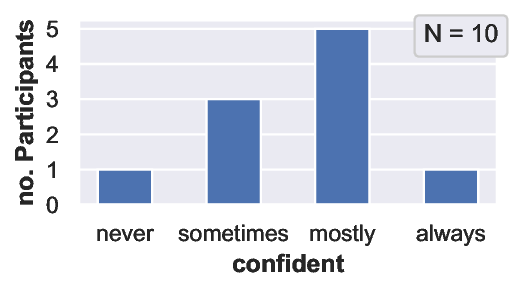}
    \caption{Distribution of confidence in specifying atomic mappings for six selected delta operations.}
    \label{fig:hist_mapping_confidence}
\end{figure}

We see two causes for the high standard deviation in the atomic mapping of delta operation (1): First, delta operation (1) was the first we asked the participants to map. Thus, we naturally expect a lower confidence, compared to the remaining five atomic mappings. Second, four participants gave evidence (code \code{00-0}{A}) for the description model of \ac{sge} to be a highly subjective matter, most generally because the mapping of a change into \acp{av} and \acp{pv} depends, for example, on individual knowledge of the person itself. However, we also asked about each participant's confidence in specifying these atomic mappings, and the majority of the participants felt mostly confident, as shown in \autoref{fig:hist_mapping_confidence}. Furthermore, compared to our own atomic mapping specifications, which are based on the understanding of Albers et al. \cite{albersModelSGESystem2022}, and shown in \autoref{tab:mapping_scenario_own_specs}, nearly all participants atomic mapping specifications match up. This indicates that, although subjectivity plays a role in the specification of atomic mappings, most of the participants have, whether knowingly or unknowingly, a similar understanding in the application of the description model of \ac{sge}.

\begin{nquote}
\textbf{Finding F1}: Even though participant subjectivity plays a role in specifying atomic mappings for our subject system, most of the participants have a similar understanding within the description model of \ac{sge}.
% Although all participants share background in the description model of \ac{sge}, they partially show high variance in its application to atomic mappings, while feeling mostly confident in their doing individually.}
\end{nquote}

\begin{table*}[h]
    \centering
    \begin{subtable}[]{0.74\linewidth}
        \centering
        \begin{tabular}{@{}lC{15mm}C{15mm}l@{}}
            \toprule
             & \multicolumn{3}{c}{\textbf{Participants}} \\ \cmidrule(lr){2-4}
             & \multicolumn{2}{c}{\textbf{Distribution*}} & \\
            \multirow{-3}{*}{\textbf{Delta Operation}} & \ac{av} & \ac{pv} & \multirow{-2}{*}{\begin{tabular}[c]{@{}l@{}}\textbf{Stdd. Dev.}\\ \textbf{($\sigma$)}\ref{footnote:standard_deviation}\end{tabular}} \\ \midrule
            (1) Add Input Port to Component & \cellcolor{lightgreen}5 & \cellcolor{lightgreen}5 & 0.5 \\ 
            (2) Add Connector & 3 & \cellcolor{lightergreen}7 & 0.458 \\
            (3) Modify Name of Component & \cellcolor{lightergreen}10 & 0 & 0.0 \\
            (4) Modify Source Port of Connector & 2 & \cellcolor{lightergreen}8 & 0.4 \\
            (5) Remove Connector & 2 & \cellcolor{lightergreen}8 & 0.4 \\
            (6) Remove Component & \cellcolor{lightergreen}7 & 3 & 0.458\\ \midrule
            \multicolumn{4}{r}{\footnotesize{*among number of participants (in total 10) who mapped to either \ac{av} or \ac{pv}}} \\ \bottomrule
        \end{tabular}
        \caption{}
        \label{tab:mapping_scenario_results}
    \end{subtable}
    \hfill
    \begin{subtable}[]{0.25\linewidth}
        \centering
        \begin{tabular}{@{}c@{}}
            \toprule
            \multirow{3}{*}{\textbf{\begin{tabular}[c]{@{}c@{}}Mapping\\Specif.\\from \resq{2}\end{tabular}}} \\
             \\
             \\[5pt] \midrule
            \ac{av} \\
            \ac{pv} \\
            \ac{av} \\
            \ac{pv} \\
            \ac{pv} \\
            \ac{av} \\ \midrule
             \\ \bottomrule
        \end{tabular}
        \caption{}
        \label{tab:mapping_scenario_own_specs}
    \end{subtable}
    \caption{Distribution and standard deviation for six selected delta operations (a), mapped to either \ac{av} or \ac{pv} (predominant one \textcolor{darkgreen}{colourised}), by in total ten participants, and (b) our own mapping from \resq{2}.}
\end{table*}

\paragraph{Interview Phase}

We aligned the interviews along our interview guide with three main questions Q10, Q20 and Q21, and inductively extracted four code categories. From the answers to Question Q10 (Plausibility, \resq{3.1}), we extracted two code categories: (1) Whether the participants think the concrete change descriptors for the presented evolution scenario are convincing (\textit{Result Convincingness}, code category \codecategory{10–0}), and (2) whether the participants approve our procedure to generate change descriptors (\textit{Procedure Approval}, code category \codecategory{10–1}). From the answers to Question Q20 (Practical Applicability, \resq{3.2}), we extracted one code category: Whether the participants think our description approach in general is supportive in model-based systems engineering (\textit{Approach Supportiveness}, code category \codecategory{20-0}). From the answers to Question Q21 (Practical Applicability, \resq{3.2}), we extracted one code category: Whether the participants see potential of our description approach to be used in automated system development processes without manual expert intervention (\textit{Automation Potential}, code category \codecategory{21-0}). Code categories, and codes together with their evidence and the number of participants who gave evidence, are presented in tables \ref{tab:codes_c10-0}, \ref{tab:codes_c10-1}, \ref{tab:codes_c20-0} and \ref{tab:codes_c21-0}.

\newcommand{\codetableheader}[2]{
    \textbf{{#2}} & \small{(\codecategory{{#1}})} & \multicolumn{2}{@{}c}{\textbf{evidence (rel.)}} & \textbf{no. particip.}
}
\newcommand{\codetablecaption}[3]{
    Extracted codes with absolute and relative evidence, and number of participants who gave evidence, for code category \codecategory{{#1}} \textit{{#2}} (\resq{{#3}}).
}

\autoref{tab:codes_c10-0} lists codes for code category \codecategory{10-0}, about whether participants think the generated change descriptors are convincing (\textit{Result Convincingness}), referring to Question Q10. We found a strong evidence (79\%) that most of the participants think that the generated change descriptors are actually convincing (code \code{10-0}{A}). In contrast, one participant mentioned that the change descriptor for the first delta \texttt{DHeatable} was not convincing, while the change descriptor for the second delta \texttt{DAS} was convincing (code \code{10-0}{B}). The participant specifically added, that he would have expected a higher share of \ac{pv}, regarding that adding a heating function to an electric mirror is, in the participant's intuition, a lot of \acp{pv}. However, in the atomic mapping specifications, the participant mapped five out of six delta operations to an \ac{av}, which resulted, for delta \texttt{DHeatable}, in a share of \acp{av} of 82\% and a share of \acp{pv} of 18\%. Finally, for code category \codecategory{10-0}, we found evidence (14\%) from a third of the participants that they had difficulties judging the generated change descriptors (code \code{10-0}{C}). As a reason, these participants mostly said that they do not know the evolution scenario in such detail to judge about the concrete numbers in the change descriptor, e.g. if 28\% \acp{av} is actually correct. In conclusion, the majority of the participants are convinced about the generated change descriptors for the given evolution scenario.

\begin{nquote}
\textbf{Finding F2}: The majority of participants state that the generated change descriptors for the given evolution scenario are convincing.
\end{nquote}

\begin{table*}[h]
    \centering
    %\begin{tabular}{@{}ll@{\hskip 0.5cm}R{24mm}@{\hskip 1.5mm}R{9mm}l@{}}
    \begin{tabular}{@{}ll@{\hskip 0.5cm}rrr@{}}
    \toprule
    \codetableheader{10-0}{Result Convincingness}\\ \midrule
    convincing & \small{(\code{10–0}{A})} & 33 & (79\%) & 9 \\
    %convincing & \small{(\code{10–1}{A})} & \cellcolor{lightergreen} 33 & \cellcolor{lightgreen} (79\%) & 9 \\
    partially convincing & \small{(\code{10–0}{B})} & 3 & (7\%) & 1 \\
    difficult to judge & \small{(\code{10–0}{C})} & 6 & (14\%) & 4 \\ \bottomrule
    %\multicolumn{5}{r}{\footnotesize{*total number participants: 12}} \\ \bottomrule
    \end{tabular}
    \caption{\codetablecaption{10–0}{Result Convincingness}{3.1}}
    \label{tab:codes_c10-0}
\end{table*}

\noindent\autoref{tab:codes_c10-1} lists codes for code category \codecategory{10-1}, about whether participants approve our procedure to generate change descriptors (\textit{Procedure Approval}), again referring to Question Q10. As described before, two participants (with 21\% evidence) did not specify atomic mappings at all, thus gave reasons for not approving our procedure to generate change descriptors (code \code{10-1}{C}). In contrast, we found evidence (46\%) from two thirds of participants that they approve the procedure in general (code \code{10-1}{A}). However, five participants (with 33\% evidence) also mentioned possible conditions to the procedure (code \code{10-1}{B}): In detail, three participants mentioned that the delta operations need to be less abstract, to be understandable by mechanical engineers, and to be usable for describing change in general. As another condition, two participants mentioned that the delta operations need to be related to a reference system, so that it is traceable if and how the change occurred before already, within the project or the company. In conclusion, most participants approve our current, or an extended procedure to generate change descriptors.

\begin{nquote}
\textbf{Finding F3}: Participants approve our procedure to generate change descriptors, but also mention conditions to the procedure to be applicable.
\end{nquote}

\begin{table*}[h]
    \centering
    \begin{tabular}{@{}ll@{\hskip 0.5cm}rrr@{}}
    \toprule
    \codetableheader{10-1}{Procedure Approval}\\ \midrule
    approved & \small{(\code{10–1}{A})} & 11 & (46\%) & 8 \\
    conditionally approved & \small{(\code{10–1}{B})} & 8 & (33\%) & 5 \\
    not approved & \small{(\code{10–1}{C})} & 5 & (21\%) & 2 \\ \bottomrule
    %\multicolumn{5}{r}{\footnotesize{*total number participants: 12}} \\ \bottomrule
    \end{tabular}
    \caption{\codetablecaption{10–1}{Procedure Approval}{3.1}}
    \label{tab:codes_c10-1}
\end{table*}

\noindent\autoref{tab:codes_c20-0} lists codes for code category \codecategory{20-0}, about whether participants think our description approach is supportive in model-based systems engineering (\textit{Approach Supportiveness}), referring to Question Q20. All twelve participants (with 31\% evidence) stated that our description approach is supportive for indicating, documenting or archiving change (code \code{20-0}{A}), meaning supportive for making change explicitly visible. Around a third of the participants mentioned two other support possibilities: first, for comparing different system development alternatives (code \code{20-0}{B}, with 6\% evidence), for example, which alternative has a higher or lower share of changed modelling elements, and hence a higher or lower development risk \cite{albersProductGenerationDevelopment2015}; second, for synthesis of change (code \code{20-0}{C}, with 4\% evidence), for example, for developing changes below an upper boundary for the proportion of \acp{pv}, hence \emph{limiting} development risk \cite{albersProductGenerationDevelopment2015}. As extracted in code \code{20-0}{D}, most of the participants (with 53\% evidence) explained seven possible extensions, to make change descriptors more supportive:

\begin{enumerate}
    \item Change descriptors could be contextualized, for example with an assessment, such as a traffic light system, or an follow-up action, such as a meeting after a change with high share of \acp{pv}. 
    \item Change descriptors could relate the described change to a reference system or, in general, a repository for already known changes. An analogue extension was mentioned for the mapping specifications already (code \code{10-2}{B}), where some participants asked to relate single delta operations to reference systems.
    \item Change descriptors could be aggregated so that they describe change on a higher level than on delta operations. For example, delta \texttt{DHeatable} of the subject system could be described with "[...] component added, function added, [and] \ac{pv} or \ac{av} in tendency" (Participant 12). We argue, however, that the inclusion of model-specific elements in a change descriptor, such as "component" from \acp{cam}, reduces independence from the model itself, making the change descriptor less versatile to be used for interdisciplinary communication.
    \item Five participants asked for explicit information on the share of \acp{cv}. In our use of the description model of \ac{sge}, this translates to the share of modelling elements which were not changed, and which is given only implicitly. 
    \item Change descriptors could indicate how distributed changes are in a model, for example whether they modify only a single element, or modify nearly all elements. 
    \item One participant mentioned that the whole description approach could be extended so that an engineer could query existing changes with dedicated in-depth questions on demand. 
    \item One participant pointed out that the "quality of the data", as given in our subject system, is "available only in the end of a project" (Participant 11), thus our description approach should also be able to cope with lower quality models. Independent of what "model quality" actually means, we argue that changes to a model, formalised with delta operations, are applicable as long as the model is syntactically correct, meaning a valid instance of its meta-model. So, even if the model still is in early development, we can generate change descriptors for every change, as long as the model is syntactically correct. 
\end{enumerate}

\noindent As last code of code category \codecategory{20-0}, three participants (6\% evidence) emphasized that our description approach serves as a foundation of technical support (\code{20-0}{E}), for managing changes in model-based systems engineering per-se. Concluding code category \codecategory{20-0}, all participants see our description approach supportive for indicating, documenting or archiving change, and also give diverse possible extensions to further enhance supportiveness.

\begin{nquote}
\textbf{Finding F4}: Participants see our description approach mostly supportive for indicating, documenting or archiving change while stating diverse extensions to enhance its supportiveness.
\end{nquote}
    
\begin{table*}[h]
    \centering
    \begin{tabular}{@{}ll@{\hskip 0.5cm}rrr@{}}
    \toprule
    \codetableheader{20-0}{Approach Supportiveness}\\ \midrule
    \begin{tabular}[t]{@{}l@{}}supp. for indicating, documenting,\\ archiving change\end{tabular} & \small{(\code{20-0}{A})} & 26 & (31\%) & 12 \\[14pt]
    %\begin{tabular}[t]{@{}l@{}}supp. for indicating, docu-\\menting, archiving change\end{tabular} & \small{(\code{20-0}{A})} & 26 & (31\%) & 12 \\[14pt]
    \begin{tabular}[t]{@{}l@{}}supp. for comparing different\\system development alternatives\end{tabular} & \small{(\code{20-0}{B})} & 5 & (6\%) & 4 \\[14pt]
    supportive for synthesis of change & \small{(\code{20-0}{C})} & 3 & (4\%) & 3 \\[2pt]
    conditionally supportive & \small{(\code{20-0}{D})} & 45 & (53\%) & 11 \\[2pt]
    as foundation for technical support & \small{(\code{20-0}{E})} & 5 & (6\%) & 3 \\ \bottomrule
    %\multicolumn{5}{r}{\footnotesize{*total number participants: 12}} \\ \bottomrule
    \end{tabular}
    \caption{\codetablecaption{20-0}{Approach Supportiveness}{3.2}}
    \label{tab:codes_c20-0}
\end{table*}

\noindent\autoref{tab:codes_c21-0} lists codes for code category \codecategory{21-0}, about whether participants see potential of our description approach to be used in automated system development processes without manual expert intervention (\textit{Automation Potential}), referring to Question Q21. Two participants (with 35\% evidence) mention use of our description approach for automation in development within constraints (code \code{21-0}{A}). For example, if the proportion of \acp{pv} of two development alternatives differs strongly from each other, then the decision for one the two development alternatives may be automated. This relates highly to the supportiveness of our description approach for comparing development alternatives in general (see code \code{20-0}{B}). Again two participants (with 18\% evidence) see automation potential of our description approach if it would additionally give verbose reasoning behind its decisions (code \code{21-0}{B}). In contrast to codes \code{21-0}{A} and \code{21-0}{B}, four participants (with 47\% evidence) could not see any use of our description approach in automated system development processes (\code{21-0}{C}), and individually gave four reasons: First, two participants generally assess the impact of automated decisions as too high. Second, one participant would make decisions not solely based on quantitative metrics, such as the share of changed modelling elements. Third, one participant misses clear project management responsibilities if decisions would be made automatically. Fourth, one participant sees human creativeness in development processes as not negligible to be replaced. In conclusion, the participants do not see clear potential of our description approach to be used in automated system development processes.

\begin{nquote}
\textbf{Finding F5}: Participants are mostly cautious with respect to an automated use of our description approach in system development processes.
\end{nquote}

\begin{table*}[h]
    \centering
    \begin{tabular}{@{}ll@{\hskip 0.8cm}rrr@{}}
    \toprule
    \codetableheader{21-0}{Automation Potential}\\ \midrule
    \begin{tabular}[t]{@{}l@{}}usable for automation\\within constraints\end{tabular} & \small{(\code{21-0}{A})} & 6 & (35\%) & 2 \\[14pt]
    \begin{tabular}[t]{@{}l@{}}conditionally usable\\for automation\end{tabular} & \small{(\code{21-0}{B})} & 3 & (18\%) & 2 \\[14pt]
    not usable for automation & \small{(\code{21-0}{C})} & 8 & (47\%) & 4 \\ \bottomrule
    %\multicolumn{5}{r}{\footnotesize{*total number participants: 12}} \\ \bottomrule
    \end{tabular}
    \caption{\codetablecaption{21-0}{Automation Potential}{3.2}}
    \label{tab:codes_c21-0}
\end{table*}

\subsubsection{Discussion}
\label{sec:eval:rq3:discussion}

Regarding \resq{3.1} on plausibility, we found evidence that the participants have a common understanding of applying the description model of \ac{sge} for specifying atomic mappings (Finding F1), that they are convinced of the generated change descriptors (Finding F2), and that they approve the procedure of our description approach (Finding F3). Thus, we conclude that generated change descriptors, and the procedure to generate them, are plausible.

Regarding \resq{3.2} on practical applicability, we found strong evidence that the participants think our description approach is supportive in model-based systems engineering (Finding F4), especially for indicating, documenting, and archiving change. In contrast to supportiveness, we found only few evidence for automation potential of our description approach (Finding F5), mostly argued by limits in its current status. Only few constructive suggestions for enhancing the automation potential were given. Thus, we conclude that our description approach is practically applicable mostly in descriptive manners, and only with uncertainties in automated system processes.
%Regarding \resq{3.2} on practical applicability, we found strong evidence that the participants think our description approach is supportive in model-based systems engineering (Finding F4), especially for indicating, documenting, and archiving change. Although participants mentioned several extension possibilities, no clear tendency to one of the extension possibilities is obvious. In contrast to supportiveness, we found only few evidence for automation potential of our description approach in its current status (Finding F5). This is mostly argued by the limits of our current description approach, meaning more context, more verbose output, and more embedding into engineering processes is necessary. Thus, we conclude that our description approach is practically applicable mostly in descriptive manners, and only with extensions in automated system processes.

Regarding \resq{3.3} on extensibility, we found strong evidence that our description approach could generally be more contextualised. Specifically, delta operations could be more embedded into concrete models, to be less abstract and more understandable by engineers (Finding F3, code \code{10-1}{B}), enhancing approval of our procedure to generate change descriptors. Change descriptors itself could be more related to reference systems (Finding F3, code \code{10-1}{B}; Finding F4, code \code{20-0}{D}), for example, to trace changes within the engineering project, and more related to risk assessments, such as traffic light systems (Finding F4, code \code{20-0}{D}), enhancing supportiveness. Furthermore, we found evidence that the share of \acp{cv} could be described explicitly, and not only implicitly (Finding F4, code \code{20-0}{D}). For several other extension possibilities mentioned, no clear tendency to one of the extension possibilities is obvious. Thus, we conclude that plausibility and practical applicability, mostly referring to supportiveness, of our description approach are open for concrete extensions, while the enhancement of automation potential could not be constructively detailed out in this user study.

%% file: 01_chapters/evaluation_threats.tex
\subsection{Threats to Validity}
\label{sec:eval:threats}

We discuss possible threats to the validity of our mixed-method evaluation along the three \acp{rq}, because aspects of validity and quality differ between quantitative and qualitative evaluation methods.
%We discuss possible threats to the validity of our evaluation along the three \acp{rq}. Specifically, for our quantitative evaluation on technical feasibility (\resq{1}), we discuss internal and external validity of our implementation. For our quantitative evaluation on technical applicability (\resq{2}), we discuss internal validity, construct validity and reproducibility of our application to the \ac{bcs} case study. For our qualitative evaluation on plausibility, practical applicability and extensibility, we discuss <quality aspects go here> of our user study, along common quality criteria found in literature <cite>.%internal validity, credibility, content validity, construct validity and transferability of our user study.

\subsubsection{RQ1: Technical Feasibility / Implementation (Quantitative)}

For our quantitative evaluation on technical feasibility (\resq{1}), we discuss internal and external validity of our implementation, the \cpscommunicator{} framework.

\paragraph{Internal Validity}

As a threat to internal validity, we identified possible errors in our generic implementation. The generic implementation reads in deltas and delta operations, processes them iteratively with provided mappings, and writes out results of mappings. However, it does not contain explicit specifications of atomic and composed mappings, which are part of the application, shown in \resq{2} on technical applicability. Thus, we assume that code complexity may be introduced in the application earliest, not in the generic implementation. Furthermore, we applied best software engineering practices, to mitigate possible errors, in our implementation.

\paragraph{External Validity}

% abbildung des dahinterliegenden procedures korrekt?

We identified one threat to external validity. While mapping specifications and informal languages of change are concern of the application, our concept is fundamentally built upon delta modelling as formal language of change. We argue that this does not limit the generalisability of our implementation, because it is possible to create a delta dialect for all models that fundamentally consist of modelling elements and relations between modelling elements. Even if tool support for delta modelling (e.g. DeltaEcore \cite{seidlDeltaEcoreAModelBasedDelta2014}) is not applicable, a delta dialect may be created manually.

\subsubsection{RQ2: Technical Application (Quantitative)}

For our quantitative evaluation on technical applicability (\resq{2}), we discuss internal validity, external validity and reproducibility of our application to the \ac{bcs} case study.

\paragraph{Internal Validity}

We identified two threats to internal validity, regarding our subject system, the \ac{bcs} case study: (1) The \ac{bcs} case study originally represents a system with variability in space (configurability), but we used it in the context of variability in time (evolution). Specifically, we took core model variants as initial model versions, and variant deltas as version deltas. However, according to Seidl \cite{seidlDeltaEcoreAModelBasedDelta2014}, both dimensions of variability can be used interchangeably, if not both at the same time. Thus, we see no limitations in the understanding of variant deltas as version deltas. (2) We used only a subset of all possible product configurations from the \ac{bcs} case study for product evolution scenarios, and only a reduced delta dialect for atomic mapping specifications. However, using all possible product configurations and the full delta dialect would not come with any further advantages for concluding the evaluation on technical applicability, as we could show technical applicability already with the subset of product configurations and the reduced delta dialect.

\paragraph{External Validity}

We identified one threat to external validity, specifically to generalisability: In the motivation of our work, we claimed that our description approach is applicable in a model-based engineering context, and also brought a running example from the \ac{sysml}/\ac{uml} world. In the \ac{bcs} case study, we then used our own \ac{emf} meta-models for \acp{cam} and \acp{smm}, and not \ac{sysml} or \ac{uml} derivatives. However, as this research question concerns technical applicability, we argue that working with real \ac{sysml} or \ac{uml} meta-models should technically be equivalent, as there exists tool support for these modelling languages based on \ac{emf}, for example Eclipse Papyrus \cite{nyamsiEclipsePapyrusFramework2025}. In conclusion, we therefore see the existence of change descriptors for our own \ac{emf}-based models as direct evidence for technical applicability.

\paragraph{Reproducibility}

For reproducibility of this evaluation, we provide our used subject system in a Zenodo package \cite{ochs_inter-disciplinary-description--changes--model-based-engineering_2026}, and the code base of the \cpscommunicator{} framework and its application in \resq{2} publicly on GitHub\footnote{\cpscommunicator{} on GitHub: \cpscomm}. Thus, the application results should be reproducible.

\subsubsection{RQ3: User Study (Qualitative)}

For our qualitative evaluation on plausibility, practical applicability and extensibility, we discuss relevant quality criteria for qualitative methods found in literature \cite{ralphEmpiricalStandardsSoftware2020}; namely internal validity, construct validity, reflexivity, credibility, and transferability.

\paragraph{Internal Validity}
% what are correlations betw evidence and no. participants? (p=0.05 sign)
%% C10-0: 0.958 (no, 0.186)
%% C10-1: ~1.0 (y)
%% C20-0: 0.899 (y, 0.038)
%% C21-0: 0.812 (no, 0.396)
We identified two threats to internal validity in our user study: (1) The selected evolution scenario, as subject system, may not be suitable for the selected participants. However, we selected this evolution scenario, as extensively described in the according section, because we assumed that participants from mechanical engineering are able understand \acp{cam} as well as the changes represented in the evolution scenarios. We further gave them an introduction on \acp{cam} and explained the evolution scenario to them in detail. (2) Data analysis, meaning extraction of relevant segments, codes and code categories from the transcribed answers of our participants, may be subjective. To mitigate subjective influence, we used multiple eye principle among the authors for data analysis.

\paragraph{Construct Validity} % why we actually measure plausibility, practc applic, extensib
We argue that we actually evaluate plausibility of change descriptors and the procedure to generate them, because we explicitly asked the participants thereof (Question Q10), because we let them explain in detail why, and because we could cover both parts, the procedure to generate change descriptors (code category \codecategory{C10-0}) as well as the resulting change descriptors (code category \codecategory{C10-1}), in our findings. We argue that we actually evaluate practical applicability of our description approach, because, again, we explicitly asked the participants thereof (Question Q20, Question Q21), and because we assume our participants have the necessary background in their research field to be able to assess practical applicability in context of the description model of \ac{sge}. We argue that we actually evaluate extensibility, because we collected mentioned extension possibilities from the participants answers in an exploratory manner.

\paragraph{Reflexivity / Interactions \& Biases}
The sessions with the participants were done by one single author (interviewer) from the field of software engineering. As our participants are from the field of mechanical engineering, few to no prior interaction between interviewer and participants happened. Furthermore, the interviewer, from software engineering, was not biased in the participants research field, mechanical engineering. For data analysis, biases towards individual participants were mitigated by first transcribing all sessions, then anonymising transcriptions, then analysing transcriptions.

\paragraph{Credibility of Participants}
% also mention low bias here
We argue that our participants are credible, because we designed the procedure of our user study the way that each participant has the ability to understand the research rationale (introduction phase), the subject system (introduction phase), the description approach (specification phase and application phase), with its procedure and its results, before the interview. Thus, we expect that the participants gave \emph{credible} answers in the interviews itself. Furthermore, we expect that participants gave \emph{honest} answers in the interviews, because of the discussed low interaction and bias between interviewer and participant.

%\paragraph{Resonance with Participants} 
% nur plausibility und pract applic, da extensib explorativer typ

\paragraph{Transferability of Findings}
We argue that our results are transferable to other participants from the research field of the description model of \ac{sge}. We further argue that our results are transferable to a study design where an other model-independent language than the description model of \ac{sge} is used, as long as the original intentions of model-independent languages are comparable, meaning description and analysis of development of cyber-physical systems \cite{albersModelSGESystem2022}. We see a mediocre transferability to other subject systems, as this depends highly on the prior knowledge of each participant, for example which other models, than \acp{cam}, the participant knows and how well he could understand the presented evolution scenario.

%% file: 01_chapters/related_work.tex
\section{Related Work}
\label{sec:related_work}

In this chapter, we present existing work from the literature, related to change description, and take both the theoretical and practical side of change description into account. Thus, we address contributions to (1) formal notions of change, to (2) semantic analysis of changes, and (3) to informal change description.

\paragraph{Formal Notions of Change}

In related work on formal notions of change, \emph{change} can be understood in various dimensions. Besides change in time, i.e. evolution, approaches also address change as unit of difference between variants in product lines \cite{clementsSoftwareProductLines2002}. This relates strongly to the original background of the notion of change we use in our description approach, Delta Modelling \cite{schaeferVariabilityModellingModelDriven2010}; and more generally, to two-dimensional variability in space (variants) and time (evolution), i.e. product line evolution. Seidl et al. \cite{seidlIntegratedManagementVariability2014} proposed \emph{Hyper Feature Models}, where features are versioned individually, together with a formal constraint language to analyse configurations for validity. Nieke et al. \cite{niekeConsistentFeatureModelDriven2022, niekeGuidingEvolutionProductline2022} proposed \emph{Temporal Feature Models}, where versions of feature models are related to each other for evolution traceability and planning. Michelon et al. \cite{michelonEvolvingSoftwareSystem2022} proposed a framework to specify and compose variants in evolving product lines, where features are versioned with revisions. Lity et al. \cite{lityHigherorderDeltaModeling2016} proposed \emph{Higher-Order Delta Modelling}, where delta models are evolved by higher-order deltas, based on addition, modification, and deletion of variant-deltas. As a combined problem and solution space model, Ananieva \cite{ananievaConsistentViewBasedManagement2022} proposed the \emph{Unified Conceptual Model}, where revisions of features and system in the problem space are mapped to implementation fragments in the solution space. However, all mentioned approaches consider modelling of changes for different purposes and modelling languages, but do not aim at describing those changes in a model-independent language.

\paragraph{Semantic Analysis of Changes}

Differencing methods extract changes between two modelling artefacts, for example the versioning tool Git\footnote{Git: \git} on line-based artefacts, on a syntactical level. But even if two modelling artefacts differ on syntactical level, they may describe the same system, thus do not differ \emph{semantically}. Semantic change analyses propose differencing procedures based on semantic rather than syntactic change, and therefore have a similar aim as our description approach, to extract semantics behind changes. For instance, semantic differencing approaches have been proposed for \ac{uml} class models with the \ac{ocl} by Rumpe et al. \cite{rumpeSemanticDifferenceAnalysis2024} and Maoz et al. \cite{maozCDDiffSemanticDifferencing2011}, for feature models by Drave et al. \cite{draveSemanticEvolutionAnalysis2019}, and for code artefacts by Jackson and Ladd \cite{jacksonSemanticDiffTool1994}. However, these approaches contribute to differencing of specific modelling artefacts, by analysing if changes concern semantics of a model itself, instead of semantically describing change, as our description approach does.

For semantically describing change, approaches have been proposed for \ac{uml} models, by Briand et al. \cite{briandImpactAnalysisChange2003}, where a change to a model is enriched by attributes, such as concerned modelling elements, and a textual representation of the change itself. Kehrer et al. \cite{kehrerRulebasedApproachSemantic2011} proposed semantic change descriptions for \ac{emf} models, with \emph{Semantic Lifting}, where patterns are dynamically detected for atomic sets of change operations to a model. Parnin and Goerg \cite{parninImprovingChangeDescriptions2008} proposed semantic change descriptions for bytecode models, where changes to bytecode are enriched by attributes, such as location of occurrence and involved data types, and then analysed with bytecode analysis techniques. Ma et al. \cite{maDesignChangeAnalysis2017} proposed semantic change descriptions for physical models of mechanical systems, where a graph structure based on ontologies is built from physical parts and their attributes, and a change to physical parts is analysed along the graph structure. More generally for ontologies, Javed et al. \cite{javedOntologyChangeManagement2013} proposed a process where change patterns are detected in order to provide semantics behind changes. However, although these approaches describe changes semantically, they are mostly limited to description languages which can not be used interchangeably nor are explicitly model-independent, which is the goal in our description approach.

% Steffen Becker, Boris Gruschko, Thomas Goldschmidt, and Heiko Koziolek. A process model and classification scheme for semi-automatic meta-model evolution. 

\paragraph{Informal Change Descriptions}

In related work on informal change descriptions, the term \emph{change} is used in context of natural language, i.e. less as technical term, what poses difficulties finding approaches comparable to the informal language of change we use in our description approach, the description model of \ac{sge} \cite{albersModelSGESystem2022}. While the field of \emph{Engineering Change Management} essentially provides methods and frameworks for handling changes in system development \cite{wrightReviewResearchEngineering1997, jarrattEngineeringChangeOverview2011}, it rather addresses business processes, which is not the scope of our work. 

The informal language of change we use in our description approach, the description model of \ac{sge}, uses a classification of reference systems, proposed by Albers et al. \cite{albersReferenceSystemModel2019}. For a more fine-grained classification of reference systems, further dimensions of variations, besides the variation type (\ac{av}, \ac{pv}, \ac{cv}), have been proposed by Martin et al. \cite{martinModelBasedApproachAnalyze2026}. The authors propose change complexity, change novelty, and prior knowledge of the reference system as further dimensions for variations. This classification is integrated into the authors risk assessment framework \cite{martinCommunicationApproachModelBased2025}, so that a higher change complexity, higher change novelty and less prior knowledge of the reference system is understood to introduce more development risk. However, the mentioned approaches consider descriptive extensions or practical uses of the description model of \ac{sge}, but do not address automated generation of change descriptions based on specified mappings from formal notions of change.

\paragraph{Summary}

In summary, existing approaches in the literature either focus on the modelling side (formal modelling languages), or on practical side (informal change descriptions) of describing changes, and do not connect both sides with the intention of advancing communication of changes between different engineering disciplines in a \emph{multi-domain} engineering environment, as our description approach does. Furthermore, the listed work is not evaluated methodologically to connect the theoretical modelling side to the practical side of describing changes, to the extent in which we evaluated our work; usually, only one side is evaluated, for example conceptual soundness on the theoretical modelling side, and usability in industry on the practical side.

%% file: 01_chapters/conclusion.tex
\section{Conclusion \& Future Work}
\label{sec:conclusion}

In this paper, we presented an approach for describing model-specific changes in model-based engineering projects with model-independent languages. By combining a formal notion of change with informal descriptions of change, we bridged the gap between technical change representations and human-interpretable notions. We realised our approach by specifying mappings for translating deltas into \emph{change descriptors}. In a mixed-methods evaluation, reflecting the translation from theory to practice, we have shown that our description approach is technically realisable and applicable, and that change descriptors are plausible and practically applicable.

In the evaluation, we have shown that our description approach is technically feasible, by implementing a generic framework, called \emph{cps-communicator}, which provides functionality to specify and apply mappings to deltas and delta operations. The \emph{cps-communicator} framework enables a versatile use, as model-independent language and mappings can be specified and then applied to concrete models with loose restrictions. Furthermore, we have quantitatively evaluated technical applicability of our description approach, by applying the \emph{cps-communicator} framework to two concrete model types, \acp{cam} and \acp{smm}, from the existing \ac{bcs} case study. Essentially, we specified 28 mappings and applied them 16092 times overall, showing technical applicability even for larger models. We have qualitatively evaluated plausibility, practical applicability, and extensibility, by conducting a user study with twelve participants with varying degrees of experience. Our findings show that most participants understand and approve our description approach (plausibility), and see a supportive use, especially in indicating, documenting, and archiving changes (practical applicability). Based on these evaluation results, we are convinced that extensions to our description approach advance its practical applicability even more.

We identified two major directions for future work from the findings of our user study. Our first idea is to better contextualise changes into the model and its environment, in which changes occur. Thus, we aim to embed change descriptors deeper in a project context, for example by taking change propagation across multiple models into account (e.g. with multi-model consistency \cite{reussnerConsistencyViewBasedDevelopment2023, spanoudakisInconsistencyManagementSoftware2001}), or by taking affected product variants of a product line into account (e.g. with delta-oriented product lines  \cite{schaeferVariabilityModellingModelDriven2010, lityHigherorderDeltaModeling2016}). Second, we aim to improve the automation potential of our description approach, by extending the \emph{cps-communicator} framework with a more user-friendly interface, and with a more accessible backend. The latter could provide technical interfaces used by automated agents, for example to automatically detect model changes (e.g. with differencing techniques \cite{kehrerRulebasedApproachSemantic2011, seidlIntegratedManagementVariability2014}) and describe them accordingly.

%% file: 01_chapters/acknowledgements.tex
This work was funded by the Deutsche Forschungsgemeinschaft (DFG, German Research Foundation) – SFB 1608 – 501798263. Thanks to our textician. Thanks to Lennart Rak.